\documentclass[conference]{IEEEtran}
\usepackage{cite}
\usepackage{amsmath,amssymb,amsfonts}
\usepackage{algorithmic}
\usepackage{algorithm}
\usepackage{graphicx}
\usepackage{textcomp}
\usepackage{xcolor}
\usepackage{fancyhdr}
\usepackage{comment}
\usepackage[hyphens]{url}
\usepackage{xspace}

\newcommand{\blue}[1]{\textcolor{black}{#1}}
\newcommand{\green}[1]{\textcolor{black}{#1}}

\usepackage{subcaption}

\newcommand{\projectname}{AccQOC\xspace}

\def\BibTeX{{\rm B\kern-.05em{\sc i\kern-.025em b}\kern-.08em
    T\kern-.1667em\lower.7ex\hbox{E}\kern-.125emX}}

\title{\vspace{-13mm}\projectname: Accelerating Quantum Optimal Control Based Pulse Generation \vspace{-4mm}} 

\author{\IEEEauthorblockN{Jinglei Cheng, Haoqing Deng and Xuehai Qian}
\IEEEauthorblockA{
\textit{University of Southern California}\\
\{chen520,haoqinde,xuehai.qian\}@usc.edu}\vspace{-11mm}}

\begin{document}
\maketitle
\pagestyle{plain}


\begin{abstract}

In the last decades, we have witnessed the rapid growth of Quantum Computing. In the current Noisy Intermediate-Scale Quantum (NISQ) era, the capability of a quantum machine is limited by the decoherence time\green{, gate fidelity} and the number of Qubits. Current quantum computing applications are far from the real ``quantum supremacy'' due to the fragile physical Qubits, which can only be entangled for a few microseconds. Recent works use quantum optimal control to reduce the latency of quantum circuits, thereby effectively increasing quantum volume. However, 
the key challenge of this technique
is the large overhead due to long compilation time. 

In this paper, we propose {\em \projectname}, a comprehensive 
static/dynamic hybrid workflow to transform gate groups
(equivalent to matrices) to pulses using 
QOC (Quantum Optimal Control) with a reasonable compilation time
budget. \projectname is composed of static pre-compilation
and accelerated dynamic compilation. 
\green{After the quantum program is mapped to the quantum circuit with our heuristic mapping algorithm considering crosstalk, we leverage static pre-compilation to generate 
pulses for the frequently used groups to eliminate the 
dynamic compilation time for them.}
The pulse is generated using QOC with binary search to 
determine the latency. 
For a new program, we use the same policy to generate 
groups, thus avoid incurring overhead for the ``covered'' groups. 
The dynamic compilation deals with ``un-covered'' groups
with accelerated pulse generation.
The key insight is that the pulse of a group 
can be generated faster based on the generated pulse
of a {\em similar} group.
We propose to reduce the compilation time by generating 
an ordered sequence of groups in which the sum of similarity
among consecutive groups in the sequence is minimized. 
We can find the sequence by constructing 
a similarity graph ---
a complete graph in which 
each vertex is a gate group and the weight of an edge
is the similarity between the two groups it connects, then
construct a Minimum Spanning Tree (MST) for SG.
\blue{With the methodology of \projectname, we reached a balanced point of compilation time and overall latency. The results show that accelerated compilation based on MST achieves $\green{9.88}\times$ compilation speedup compared to the standard compilation of each group while maintaining an average $2.43\times$ latency reduction compared with gate-based compilation.}

\end{abstract}

\section{Introduction}
\label{intro}

The idea of Quantum Computer \cite{aharonov1999fault} has been proposed for decades. In recent years, we have witnessed many breakthroughs in building a quantum machine \cite{boixo2018characterizing,farhi2016quantum,kelly2015state}. IBM \cite{knight2017ibm} and Google \cite{hsu2018ces} have demonstrated working quantum machine with 50 bits and 72 bits. Google claims to have built the first quantum computer that can carry out calculations beyond the ability of today’s most powerful supercomputers \cite{arute2019quantum}. This indicates that we have entered the era of Noisy Intermediate-Scale Quantum(NISQ) \cite{preskill2018quantum}, where we expect to see a quantum machine with hundreds or thousands of Qubits outperforming classical supercomputers in the coming decades. However, the Qubits inside a NISQ device are far from perfect. First, the connection between Qubits is sparse. Second, the operations of Qubits are vulnerable to errors due to insufficient decoherence time. The sparsity of Qubits means that 2-bit operations are not supported for all Qubit pairs. Furthermore, the limited number of Qubits makes it unrealistic to implement quantum error correction code(ECC) \cite{bennett1996mixed,o2017quantum,suchara2013comparing}, which would need thousands of Qubits. Therefore, current quantum computers suffer from high gate error rates. Consequently, large-scale programs like Shor algorithm \cite{shor1999polynomial} or Grover Search algorithm \cite{grover1996fast} could not be implemented on NISQ devices.

\green{In the NISQ era, the capacity of a quantum machine is limited by its 
{\em quantum volume} --- determined by the number of Qubits and the decoherence time
--- and gate error.} 
For current IBM's quantum devices, 
the longest decoherence time achieved is less than one hundred microseconds \cite{ibm_device}. 
Thus, the reduction of latency is critical to NISQ devices for two reasons. 
First, we could run larger programs within the same decoherence time on a given quantum machine.
\green{Second, for the current quantum programs, 
the coherence error can decrease substantially by latency reduction\green{~\cite{shi2019optimized,nielsen2002quantum}}. 
Based on our calculations in Section~\ref{error_calculate},
the coherence error is {\em comparable} to gate error. 
Thus, the fidelity, affected by both coherence and gate error, 
can be improved by latency reduction. }

The current compilation for quantum system is gate-based \cite{qiskit} --- quantum programs are first compiled into specific 1-bit or 2-bit operations that the quantum machine supports. This step is often referred to as synthesis \cite{shende2006synthesis}. These hardware-supported gates will be further translated into corresponding electrical signals, often referred to as {\em pulses}. The gate-based compilation method requires little compilation time but leads to longer latency, {because the synthesis turns the desired unitary matrix into multiple instructions whose corresponding pulses will be concatenated to reach the target function. } 

To mitigate these drawbacks, physicists proposed quantum optimal control (QOC) \cite{glaser2015training}, which could directly compile quantum state transfer or a functional unitary matrix
{that emulates the operation performed by a group of gates} into control pulses. There exist multiple algorithms developed for quantum optimal control, such as gradient ascent pulse engineering (GRAPE) \cite{de2011second} and Krotov algorithms \cite{krotov1995global}, most of which are based on gradient methods. The pulses control the unitary matrix that represents the function of a quantum circuit in a systematic manner. The evolving path towards the final states could be divided into small steps. By optimizing these steps through analytical or numerical algorithms, the quantum optimal control can generate a sequence of control pulses that could approximate the target matrix \cite{glaser2015training}.

Quantum optimal control could effectively reduce the latency of groups of quantum gates. Since the size of the unitary matrix that represents the function of quantum gates does not increase with the number of gates, the latency of pulses generated by QOC based compilation does not scale with the number of gates as much as the gate-based compilation does. Recent work \cite{shi2019optimized} has utilized this method to reduce the latency of the aggregated gates. Together with proper gate scheduling, it could bring considerable latency reduction. 

\blue{On the other side, this method has the inevitable drawback of 
requiring enormous computational resources. Currently GRAPE only supports 10 Qubit for very simple operations (such as 10 concurrent single-bit operations). }
For groups composed of more than 5 bits, we observe that it could take several hours to generate the corresponding pulses even for a simple Qubit model. As a result, it is impossible to compile a quantum program with hundreds of thousands of gates within a day. To mitigate this problem,   \cite{leung2017speedup} uses automatic differentiation with the GPU to accelerate the quantum optimal control. 
\blue{However, such acceleration method only achieves significant speedup with utilization of GPU when more than 10 Qubits are involved. The acceleration for small groups is limited.}
\cite{shi2019optimized} also points out the large compilation overhead caused by QOC.  

\green{Recent work \cite{gokhale2019partial} addresses this 
problem for quantum variational algorithms such as VQE and QAOA
with partial compilation. 
The variational quantum programs are executed iteratively,
the parameters (mostly rotation angle) of the current iteration 
are determined by the output distribution of the previous iteration. 
For this specific type of algorithms, pre-computation can be applied
to obtain the proper hyperparameters for compilation
that can accelerate pulse generation of groups with 
different rotation angles.
This method reduces the compilation time 
of variational algorithms thanks to its iterative nature:
for a given group, it is ``parameterized'' with only the rotation
angles for different iterations, and the selected hyparparameters
can accelerate the pulse generation of all concrete groups with 
rotation angles determined. 
To ease the discussion, we call the concrete groups 
determined by the same parameterized group as 
a {\em group family}.
This methodology does {\em not} work for
non-variational algorithms such as Shor algorithm~\cite{shor1999polynomial}
for the following reason. 
For these algorithms, the program does not change during execution. 
Therefore, the execution does not involve groups with changing parameters. The gates implementing the whole 
algorithm can be only decomposed into {\em static groups}
which {\em do not belong to the same group family}.
If we use the method of \cite{gokhale2019partial},
then the hyparparameters of {\em each} static group need
to be generated {\em but not reused}. 
Clearly, \cite{gokhale2019partial} does not help the pulse
generation of static groups.
Moreover, the recent works that optimize the QOC based compilation do not consider its scalability, making it not realistic to compile large quantum programs.}

\green{As the first attempt to accelerate the pulse generation of static groups,} we propose {\em \projectname}, a comprehensive 
static/dynamic hybrid workflow to transform gate groups
(equivalent to matrices) to pulses using 
QOC with reasonable compilation time
budget. \projectname is composed of static pre-compilation
and accelerated dynamic compilation. 
\green{First, the quantum program is mapped to the quantum computer with our heuristic mapping algorithm. Our mapping algorithm takes into consideration the cross-talk effect. We aim to increase fidelity through the mitigation of cross-talk effect.}
Then,
we leverage pre-compilation to generate 
pulses for a category of groups to eliminate the 
dynamic compilation time for them.
To get this category of groups, we perform profiling on a randomly selected set
of programs with certain grouping policy.
The corresponding pulses are then generated using QOC with binary search to 
determine the shortest possible latency. 
Given a target program, it is first decomposed with the same
policy used in pre-compilation. For the gate groups of which 
the pulse is available, there is no compilation cost. 
We only need to focus on the groups that are not 
``covered''. \blue{When pulses of all groups are generated, they are concatenated together to determine the overall latency of the target quantum program. Since different grouping policy will lead to different overall latency, we performed six different policies on some sample programs and pick the policy of best performance.}

The second component of \projectname deals with the 
dynamic compilation of ``un-covered'' groups
with accelerated pulse generation.
In fact, the technique 
applies to both those ``un-covered'' groups as well as the static pre-compilation
(but it is a one time cost). 
The key insight is that the pulse of a group 
can be generated faster based on the generated pulse
of a {\em similar} group.
Thus, we can reduce the compilation time by generating 
an ordered sequence of groups in which {\em the sum of similarity
among consecutive groups in the sequence is minimized}. 
We can find the sequence by 
1) constructing {\em similarity graph or SG} ---
a complete graph in which 
each vertex is a gate group and the weight of an edge
is the similarity between the two groups it connects; and
2) constructing a Minimum Spanning Tree (MST) for SG.
We will include the identity matrix (recall that 
a group corresponds to a matrix) as a vertex, if 
no group is similar enough, the compilation will start from 
the pulse of identity matrix.

With this workflow, we are able to reach a balance between compilation time and latency reduction. By using ``map2b4l" strategy described in section \ref{GP}, We achieve an average of 2.43$\times$ reduction of overall latency and \green{9.88}$\times$ reduction of compilation time.
\green{\projectname is more general than the method in~\cite{gokhale2019partial}, and can handle the parameterized
groups and static groups in the same manner. 
Specifically, for the variational algorithms, 
\projectname will treat the groups with different rotation angles
simply as different static groups and accelerate the pulse generation
by keeping previously generated pulses and selecting the 
most similar group's pulse as the initial condition.
Thus, \projectname can generate new groups with arbitrary 
rotation angles --- our method does not use it as a parameter. 
As explained before, the fidelity is affected by both 
coherence error and gate error, the reduction of latency by 
QOC mainly affects coherence error. In Section~\ref{error_calculate},
we demonstrate with calculation 
that the two sources of errors are comparable.
Thus, reducing latency is an important way to improve fidelity.}

The rest of the paper is organized as follows: section  \ref{back} gives the basic background of Quantum Computing and states our main motivation; section  \ref{flow} presents an overview of our methodology; section  \ref{gen} describes in detail how pre-compilation works to reduce overall latency; section  \ref{at} describes further optimization to reduce pre-compilation time; section  \ref{eva} shows our results of pre-compilation and further optimization; we relate previous work in section  \ref{re} and summarize our work in section  \ref{con}.

\section{Background}
\label{back}
This section presents the necessary background of quantum computing and quantum optimal control.

\subsection{Basics of Quantum Bit}

The essential difference between quantum computing and classical computing lies in the property of Quantum Bit (Qubits) \cite{nielsen2002quantum}. A Qubit has an infinite number of states which are different superposition of logical states 0 and 1, rather than only two logical states as in classical bits. The state of one Qubit could be generally represented as $|\psi\rangle = \alpha|0\rangle + \beta |1\rangle$, where a and b are complex numbers satisfying ``$|\alpha|^2+ |\beta|^2 = 1$". 
The matrix that represents the state of this one bit is 
$
|\psi\rangle =
\begin{pmatrix}
\alpha \\
\beta
\end{pmatrix}
$.
When this Qubit is measured on a basis of 0/1, the quantum states collapse, and the probability of measured results being 0 and 1 is $|\alpha|^2$ and $|\beta|^2$ respectively. Similarly a quantum system of two Qubits have four orthogonal basis, and the quantum state could be represented as ``$|\psi\rangle = \alpha|00\rangle + \beta |01\rangle + \gamma|10\rangle + \delta|11\rangle$". When measured, the quantum state would have probabilities of $|\alpha|^2$, $|\beta|^2$, $|\gamma|^2$ and $|\delta|^2$ being 00,01,10,11 respectively. 
Therefore, a quantum system with N Qubits would have $2^N$ quantum states and needs $2^N$ complex parameters to describe. The exponential feature of quantum states gives quantum computers the ability to solve problems that are intractable to classical computers. It also leads to the fundamental 
difficulty in simulating a quantum system.

\subsection{Basics of Quantum Computing}
\begin{figure}[t]
\centering
\vspace{-2mm}
\includegraphics[width=0.40\textwidth]{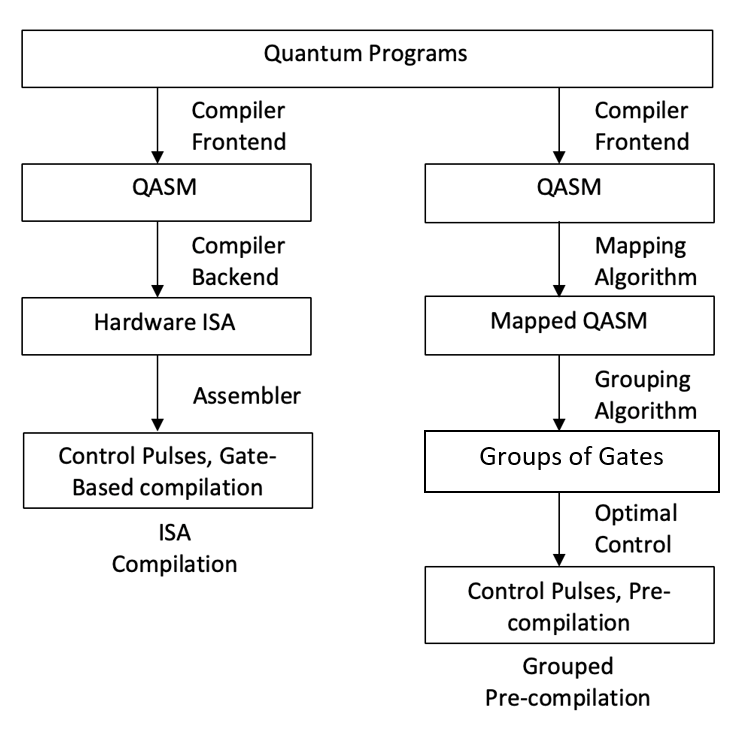}
\vspace{-4mm}
\caption{Comparison between two compilation methodologies. The left describes classic gate-based compilation approach \cite{shi2019optimized}. Our approach on the right utilizes pre-compilation.}
\vspace{-8mm}
\label{fig:label}
\end{figure}
The basic element of a quantum program is the quantum gate. These gates have different functions and could be represented by different unitary matrices. An N-bit quantum gate could be represented by a $2^N \times 2^N$ matrix. The function of multiple gates can be computed by simply multiplying matrices of the individual gate, respectively. For instance, the state transfer function of a 10-bit quantum system with 5 gates could be computed by multiplying 5 matrices of size $1024\times1024$.

In the control flow of running a quantum program, the quantum algorithm is first synthesized into several discrete basic quantum logical gates. However, these gates may not match the basic gates supported directly on quantum machines. 
For example, a Toffoli gate is commonly used in quantum algorithms. However, it could not be supported on quantum hardware \cite{fedorov2012implementation}. So it will first be decomposed into smaller gates, then translated to control pulses.
\begin{figure}[h!]
\centering
\vspace{-4mm}
\includegraphics[width=0.48\textwidth]{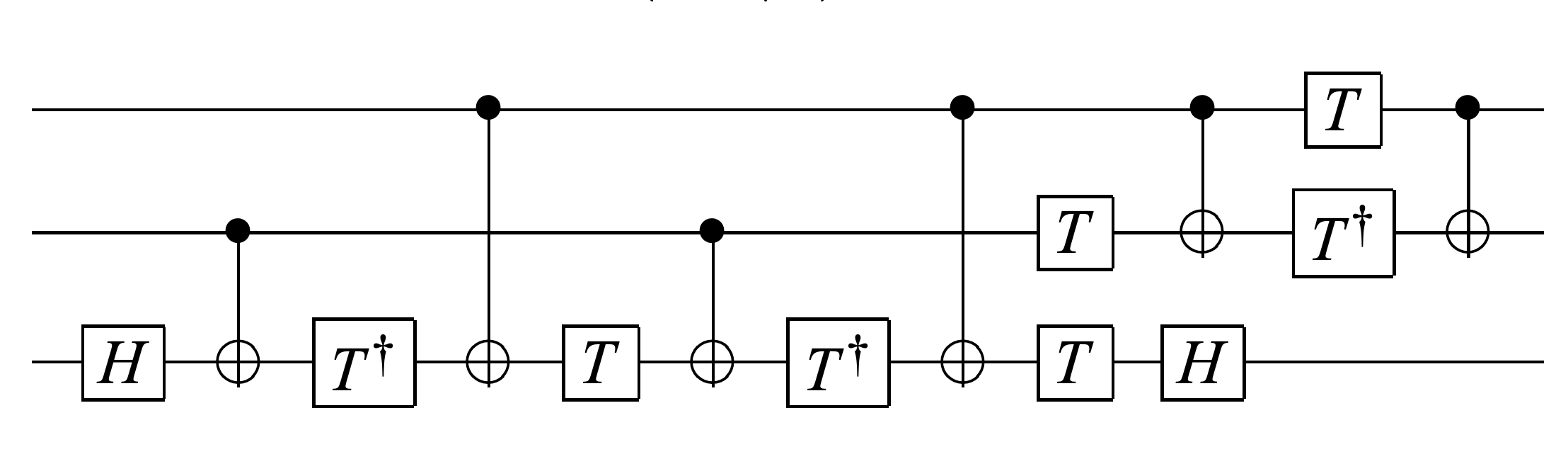}
\caption{The Toffli gate often used in quantum programs is not directly supported by a quantum computer, thus it must be decomposed into basic gates that are compatible with quantum hardware.}
\vspace{-4mm}
\label{fig:label}
\end{figure}

The second step is to map Qubits to physical bits residing on the actual machine. Since the quantum program usually can not be directly executed on certain quantum machines due to the sparse connection between quantum bits, swap gates are needed to be inserted to make the quantum program compatible with the hardware. This step could be referred to as the Qubit mapping problem. The mapping algorithms have been discussed in detail \cite{li2019tackling, mapping,wille2019mapping}. Since it is essentially an NP problem, the mapping algorithm remains possible to be further optimized. 

After the device-dependent program is generated, the gates need to be scheduled and translated to pulses, then they can be executed on quantum machines. Finally the program is run multiple times, and the result is the distribution of the output.

\subsection{Gate-based Compilation}

\begin{figure}[h!]
\centering
\includegraphics[width=0.40\textwidth]{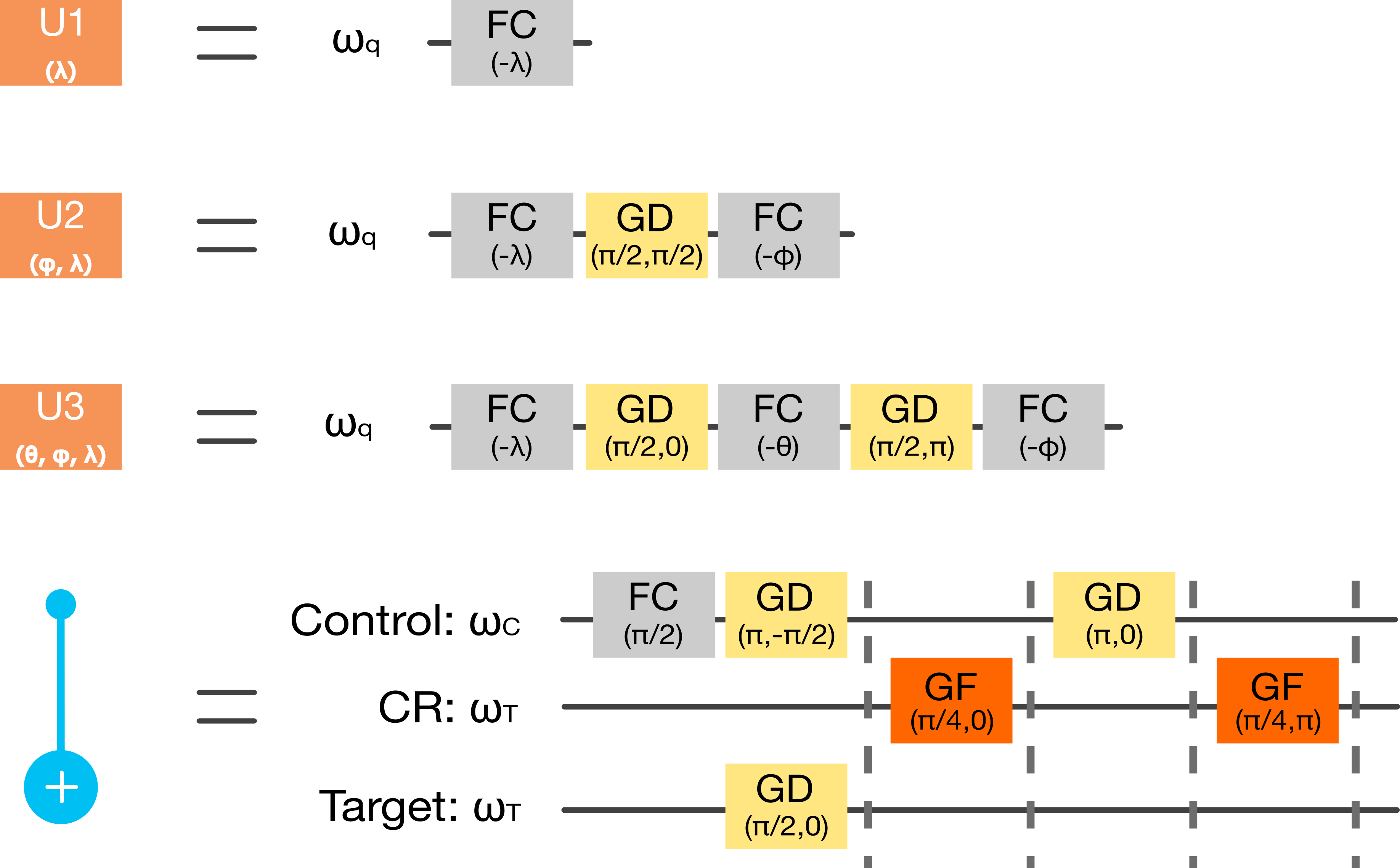}
\caption{Each gate corresponds to a pulse \cite{qiskit}. One-by-one concatenation of all pulses realize the function of many gates.}
\vspace{-4mm}
\label{gcb}
\end{figure}
The method to translate gates to pulses is gate-based compilation. This method uses a gate-pulse look-up table to find the corresponding pulse for the gate, then concatenates all pulses together to form the final pulses for the whole program. For example, current IBM quantum machines support u1,u2,u3, and CNOT gates, which contain rotation operations and common 2-bit gate. Their corresponding pulses are also given in Figure  \ref{gcb}. The category of hardware-supported gates will decide how quantum programs are decomposed. There exist mismatches between basic gates of quantum programs and basic gates supported by hardware. The Toffoli gate mentioned above is a perfect example. The 3-bit gate is not directly supported by current quantum machines. So it must first be decomposed into 15 basic gates, which are then directly translated into corresponding pulses. For gates with parameters, like rotation gates, different rotation angles determine the strength and the length of a certain pulse. 
Gate-based compilation requires minimum compilation time, but it is inefficient compared with QOC based compilation in terms of pulse latency.

\subsection{Quantum Optimal Control and GRAPE}

Quantum optimal control \cite{glaser2015training} takes advantage of a feature of quantum machines --- states of the Qubits could be manipulated with the unique and time-dependent Hamiltonian matrix \cite{leung2017speedup,peirce1988optimal,werschnik2007quantum}. Among many algorithms that are based on gradient descent, we choose one implementation of GRAPE \cite{khaneja2005optimal} as our tool of quantum optimal control. To approach the target unitary matrix, the tool uses a numerical method to evolve from initial condition to the final state. The control pulses are divided into small steps. By gradually modifying theses small steps with enough iterations, the output fidelity will increase to an acceptable value. The procedure is similar to the training algorithm of neural networks. The cost functions
(typically fidelity) could be chosen manually to emphasize the expected characteristics of output pulses. Since the path from an initial condition to target states is achieved through evolution, it is not unique. How to find the optimal control for a given target unitary matrix is still an open problem.

\subsection{\green{Fidelity: Coherence vs. Gate Error}}
\label{error_calculate}

\green{In this section, we analyze the relative importance of 
coherence error and gate error for different technologies. 
For superconducting qubits, the ratio of gate times to decoherence times is relatively high, so even after a few gates, we do experience the exponential decay due to decoherence. 
For example, the average relaxation and coherence times for Melbourne qubits are $T1 = 57.35\mu s$ and $T2 = 61.82\mu s $\cite{ibm_device} , the time needed for a CX gate is approximately $974.9 ns$\cite{ibmdesign}.
The error caused by decoherence in the $974.9 ns$ is $1-e^{-0.9749/57.35}=1.69\times 10^{-2}$. Such error is
in the same order and comparable with the average CX gate error of $2.46\times 10^{-2}$. 
The calculation shows that latency reduction can indeed 
improve overall fidelity for superconducting quantum computers 
where error caused by qubit's decoherence is relatively high.
For trapped ion system, decoherence times are extremely long relative to gate times, so the reduced latency may not improve
fidelity as much in this scenario, and just leads to faster time-to-solution.
Moreover, \cite{shi2019optimized} shows that 
shorter pulses generated by QOC have simpler shape 
than those generated by gate-based concatenation
and are easier to implement. }


\subsection{Cross-talk}

Cross-talk is an important source of noise in quantum computers. When instructions are executed in parallel, the cross-talk effect will substantially reduce the fidelity of these instructions \cite{lienhard2019microwave}. \green{Cross-talk comes from leakage of control signals, and the control signals of one instruction could interact with the control signals of parallel instructions.} The strength of the interaction is affected by the physical distance of bits these parallel instructions operate on. \green{Crosstalk noise is prevalent across many of the leading qubits such as superconducting~\cite{} and trapped ion qubits\cite{ct1,ct2}. }

\subsection{Motivation}

\begin{figure}[htb]
\begin{subfigure}{0.24\textwidth}
\includegraphics[width=0.9\linewidth]{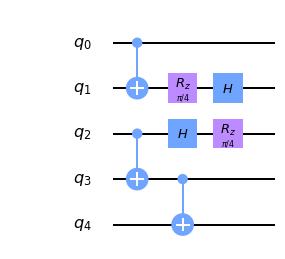} 
\caption{Group with rotation $\theta_{1}=\pi/4$}
\label{group1}
\end{subfigure}
\begin{subfigure}{0.24\textwidth}
\includegraphics[width=0.9\linewidth]{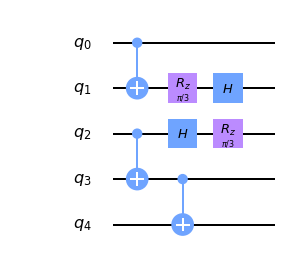}
\caption{Group with rotation $\theta_{2}=\pi/3$}
\label{group2}
\end{subfigure}

\quad

\begin{subfigure}{0.24\textwidth}
\includegraphics[width=0.9\linewidth]{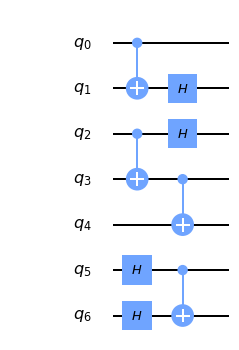}
\caption{Group with many qubits}
\label{group3}
\end{subfigure}
\begin{subfigure}{0.24\textwidth}
\includegraphics[width=0.9\linewidth]{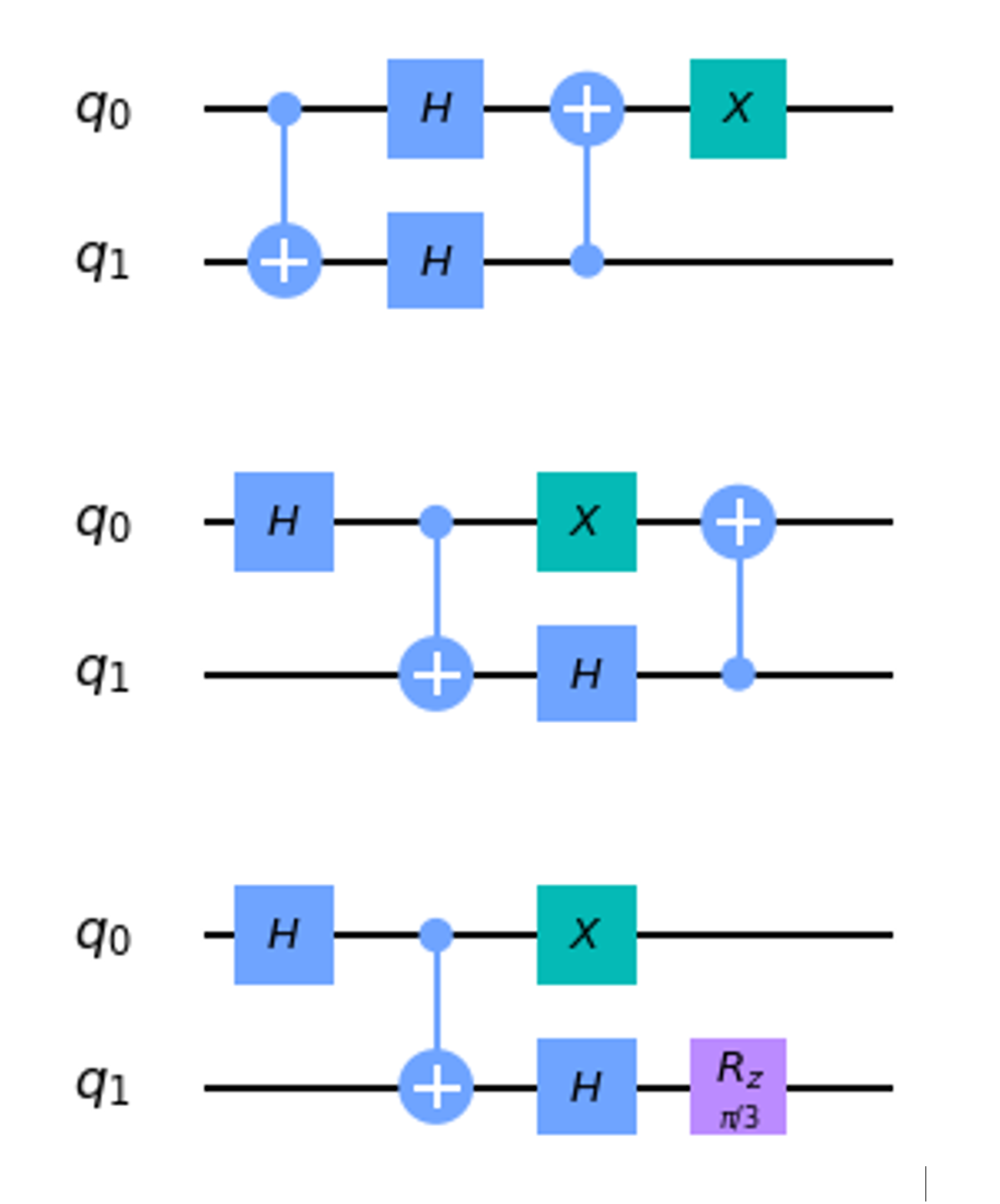}
\caption{Typical group in our paper}
\label{group4}
\end{subfigure}
 
\caption{\green{Groups generated by our method compared to recent work\cite{shi2019optimized,gokhale2019partial}. We limit our group size to ensure fast compilation and good coverage. The groups generated by \cite{shi2019optimized} could be too large and not quite scalable. The grouping algorithm proposed by \cite{gokhale2019partial} can only benefit variational algortihm.}}
\vspace{-6mm}
\label{groups}
\end{figure}

\green{Two recent works~\cite{shi2019optimized,gokhale2019partial} are
also closely related to QOC and both have the notion of grouping. 
Here we elaborate the differences by group examples shown in 
Figure~\ref{groups}.
Figure~\ref{group1} and ~\ref{group2} show 
two groups in the same ``group family'' (defined in Section~\ref{intro})
for variational algorithms, we can see that the group structures 
are the same and the only difference is the rotation angles 
($\theta_{1}=\pi/4$ vs. $\theta_{2}=\pi/3$).
~\cite{gokhale2019partial} generates the hyparparameters
(e.g. learning rate) by pre-computation, which can accelerate the pulse generation
of groups in the same family.
In comparison, the goal of \projectname is to accelerate the 
pulse generation for {\em static groups} after decomposition
as shown in Figure~\ref{group4}.
We can see that these groups can be completely different, 
instead of just having different rotation angles. 
Therefore, they clearly belong to the different group family 
with different hyparparameters. 
Also, each group just needs to be compiled {\em once}.
This explains why the method of ~\cite{gokhale2019partial} 
does not apply to non-variational algorithms.}

\green{Compared to the groups of ~\cite{shi2019optimized} shown 
in Figure~\ref{group3}, the groups of \projectname is much smaller. 
This is due to the different goals. 
The goal of ~\cite{shi2019optimized} is to achieve higher 
parallelism and minimize the latency of pulses. Specifically, it finds the commutative gates that
provide more flexibility in scheduling, i.e., execute the gates
in the alternative order. Then, Commutativity-aware Logical
Scheduling (CLS) attempts to schedule many gates
to achieve high parallelism and thus reduce latency. 
As the result, the group size tends to be large. \green{Without manually limiting the number of qubits or layers,  the aggregation methodology discussed in \cite{shi2019optimized} would generate groups with up to 10 qubits. It is costly to generate pulses for groups of this size.}
The goal of \projectname is to accelerate pulse generation with
the central idea of pre-compilation and group similarity. 
To ensure a good coverage, the group size needs to be relatively 
small (such as the examples in Figure~\ref{group4}).
Moreover, a large group size will lead to many possible groups (matrices)
of that size, making it hard to take advantage of group similarity. }




\begin{figure}[t]
\centering
\includegraphics[width=0.35\textwidth]{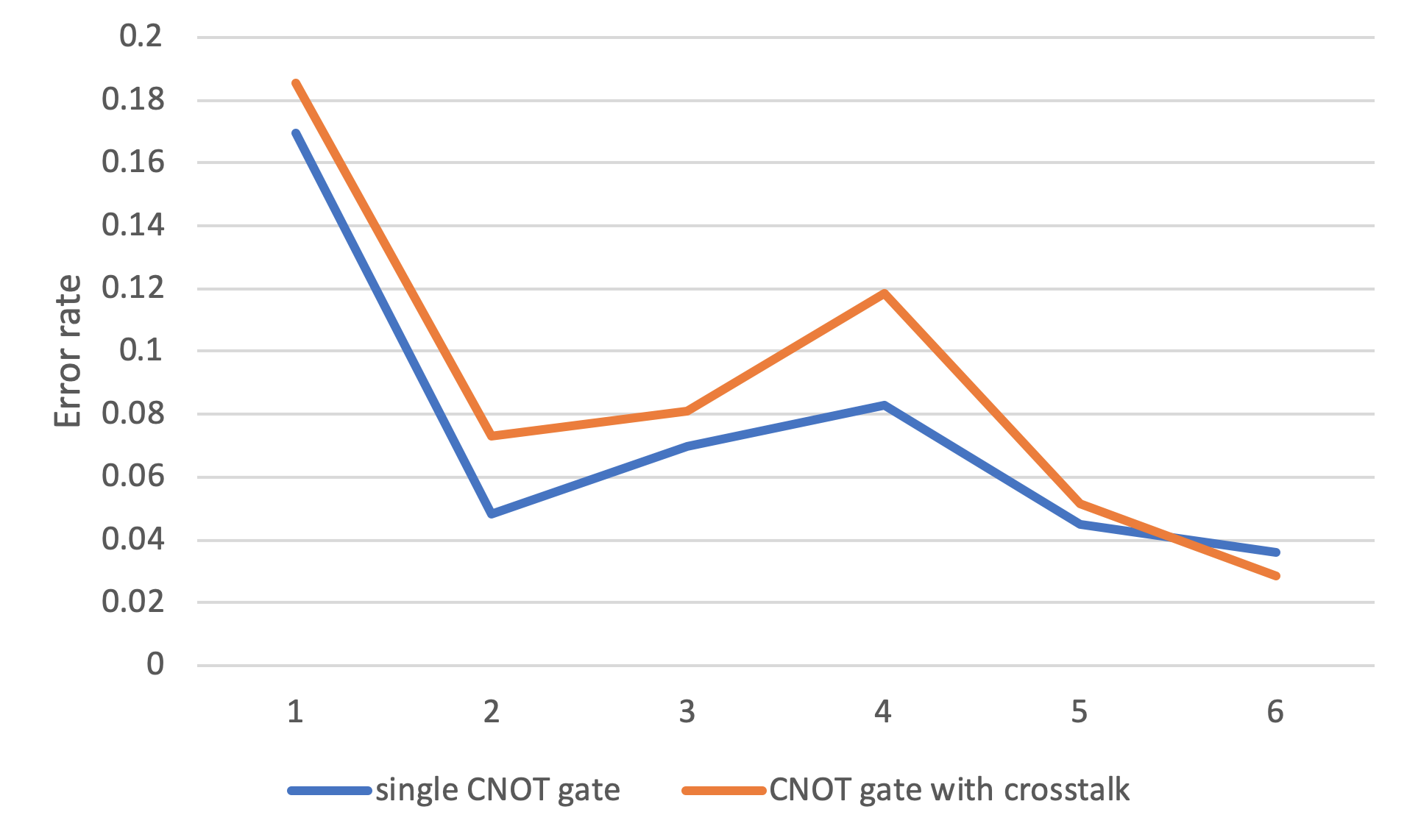}
\vspace{-3mm}
\caption{Crosstalk and Error Rate. }
\vspace{-6mm}
\label{ctef}
\end{figure}

\green{From the discussion, we see that QOC can considerably reduce latency, but with a huge computation overhead.
The prior work~\cite{gokhale2019partial} addresses this problem for 
variational algorithms, but the long compilation time 
for non-variational algorithms is still an {\em open problem}.
This paper makes the first attempt to accelerate {\em pulse 
generation for static groups} after decomposition based on certain policy.}
 \blue{Specifically, our solution leverages 
 pre-compilation and gate group similarity to accelerate the pulse generation with QOC. 
 With pulses generated for groups, 
 the latency of a given program can be greatly reduced.
 Nevertheless, our solution is a step toward faster pulse generation but has not yet fully addressed the problem of QOC's large overhead of compilation time, which scales with the size of the input quantum program.} 

Moreover, considering the two extreme policies of dividing the program into many one-gate groups and into one many-gates large group: the latter could achieve substantially more reduction in latency than the former with huge compilation overhead. However, the two extremes tell us little about what is in between. In essence, our solution finds a balanced point between the compilation time and the latency reduction. 

\section{\projectname Overview}
\label{flow}
\begin{figure}[t]
\centering
\includegraphics[width=0.50\textwidth]{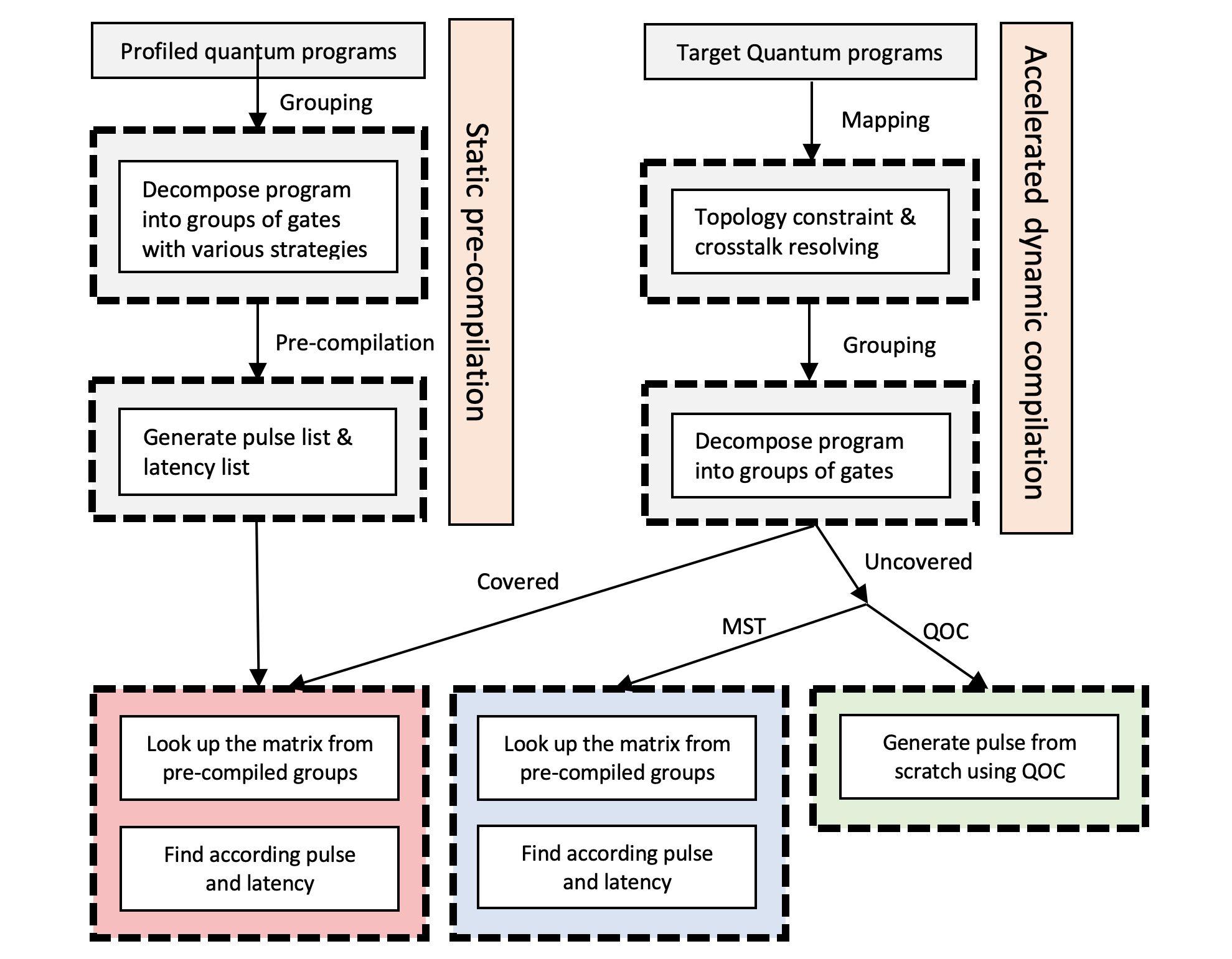}
\caption{Our method utilizes pre-compilation to generate a group list and a pulse list for future reuse, so that they do not need to undergo the time-consuming compilation process again. For groups that are ``uncovered'' by pre-compilation, we construct the MST to accelerate the computation of their pulses.}
\vspace{-6mm}
\label{overview}
\end{figure}

Figure  \ref{overview} explains the overview of \projectname
and highlights the
main contributions 
compared with standard gate-based compilation. 
Our framework shares the same part of compilation front-end with gate-based compilation. But for the part of the translation from quantum gates to control pulses, we use pre-compilation method explained in Section  \ref{gen} rather than a gate-based compilation. 
The first component is the {\em static pre-compilation}.
A category of groups of gates, whose size is determined by parameters of different decomposition policies, is first generated through randomly profiling a set of quantum programs inside our benchmark.
We use GRAPE \cite{peirce1988optimal} to generate the pulse list and latency list from the category of groups, respectively.
The pulse generation is based on QOC with binary search 
to determine the latency. 
To generate the overall latency,
dynamic programming is used solve 
the gate dependency issue and concatenate pulses of all groups together. 
\blue{Different decomposition policies are compared against each other in terms of their corresponding overall latency, and we choose the one that produces the lowest latency.}
Based on this policy, the pulses of the most frequent groups will be re-generated with different parameter settings to gain better latency.
In summary, this step pre-compiles pulses for a set of 
gate groups.

The second component is the {\em accelerated dynamic compilation}.
Given a new quantum program, we first decompose it into 
groups using the chose policy above. Some groups will be 
``covered'' by the pre-compiled pulses, and they can be 
directly used. 
For the ``un-covered'' groups, we generate an
ordered sequence from the complete similarity graph (ST)
with the similarity between a pair of 
groups as the weight of each edge. 
This sequence minimizes {\em the sum of similarity
among consecutive groups in the sequence}
and can be generated by computing the 
Minimum Spanning Tree (MST) of SG. 
After the pulse of all groups are obtained, 
the pulses for the whole program is concatenated 
in a similar manner as in pre-compilation 
based on DAG with dynamic programming. 

Overall, \projectname accelerates the
pulse generation based on QOC with static pre-compilation
and accelerates dynamic compiling due to similarity between groups.

\section{Static Pre-Compilation}
\label{gen}

\subsection{Cross-talk Consideration}
\label{crosstalk}

\blue{We take the cross-talk effect into account in the mapping process, because pulses are generated from hardware-dependent quantum programs after swap insertion.}
When inserting swap gates to make the quantum program compatible with the hardware topology, we utilize the mapping tool developed by  \cite{mapping}. This method uses an A* search with a heuristic function to find the mapping. 

We extend the heuristic function to take the cross-talk effect into consideration. The extended heuristic function is:

$$h(\sigma_{i}^{j}) = \sum_{g\in l_{i}}h(g,\sigma_{i}^{j}) + 
\sum_{g_{m},g_{n} \in l_{i}}I_{g_{m},g_{n}}$$

Where the $h(\sigma_{i}^{j})$ is the function that represents the heuristic cost of a layer $l_{i}$'s mapping $\sigma_{i}^{j}$. And the function $h(g,\sigma_{i}^{j})$ calculates the physical distance between the two Qubits of gate $g$. The extension we add is the indicator function of $g_{m}$ and $g_{n}$ which will be 1 if these two gates are too close. 

\begin{figure}[htb]
\centering
\vspace{-4mm}
\includegraphics[width=0.49\textwidth]{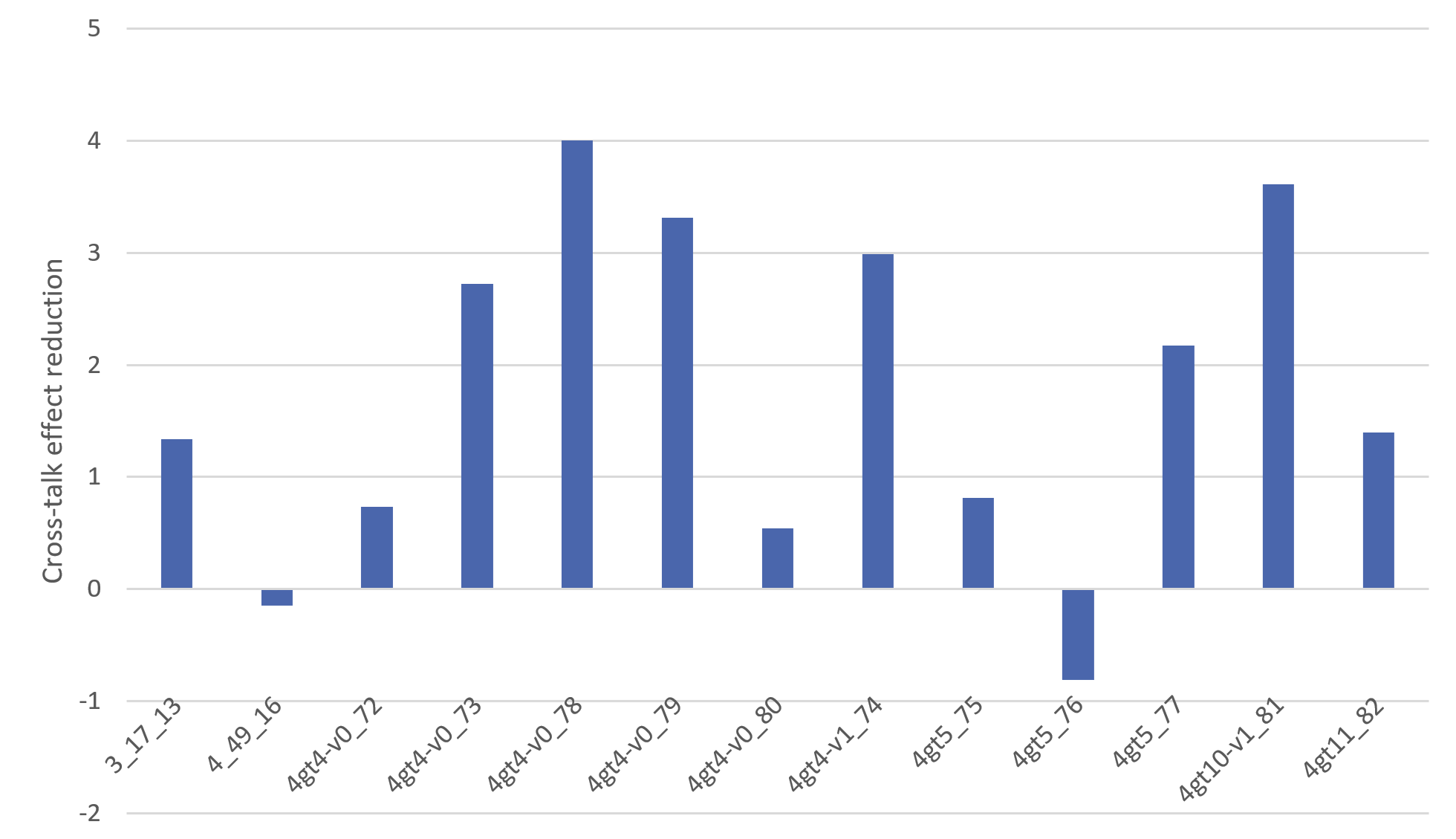}
\caption{\green{The effect of cross-talk effect is quantified by the sum of occurrences of nearby pairs of CNOT gates. After the heuristic takes it into consideration, the cross-talk effect deceases for most of the tested quantum programs.}}
\vspace{-4mm}
\label{IF}
\end{figure}

\blue{Figure  \ref{ctef} shows the cross-talk effect of real quantum computer IBM Q Melbourne.
The x-axis indicates different Qubit pairs and 
y-axis shows error rate.
Different Qubit pairs have differnet error rate. 
In general, the lower the curve, the higher 
fidelity is achieved. 
The lower curve shows the results for the single 
CNOT gate without cross-talk. 
The higher curve represents the fidelity of a CNOT gate operating on the same pair of Qubits but affected by a nearby CNOT gate. We see that these six Qubits pairs suffer from average 20\% higher error rate due to the effect of cross-talk.}
To evaluate the effects of cross-talk consideration
as a part of the mapping algorithm,
we quantify the total cross-talk effect as the sum of occurrences of close CNOT pairs in each layer. \green{This metric is adopted from the paper \cite{murali2020software}. The qubits are dispersively coupled on quantum devices, which means that the cross-talk effect is much stronger on closer qubits. This metric also makes characterizing crosstalk more efficient\cite{murali2020software}.}  \blue{As the results shown in Figure \ref{IF} in
Section \ref{cross}}, we are able to achieve \blue{17.6\%} reduction of the total cross-talk effect. 

\subsection{Grouping policy}
\label{GP}
We try several policies of generating gate groups 
based on different values of n in the 2bnl cataloging system, where 2 represents the maximum number of Qubits in a gate group and n the number of layers. We limit the maximum number of Qubits to be 2 because a group composed of more than 2 Qubits takes too much time to train with QOC, which violates our goal of reducing compilation time. We choose 2b2l, 2b3l, 2b4l as candidate policies, for their relatively small size makes it realistic to profile.

There are two ways of dealing with swap operations generated during the mapping process: 
treating swap as an independent operation or decomposing swap into three CNOT gates. Since different quantum machines may implement swap operations in both ways, we experiment with both methods, and differentiate them as ``map'' and ``swap''. For each method, we experiment with the 3 candidate policies mentioned above.

\begin{table}[h!]
\centering
\begin{tabular}{c|c|c}

Different policies & \# Qubits & \# Layers \\ \hline
swap / map           & 2         & 2         \\ \hline
swap / map           & 2         & 3         \\ \hline
swap / map           & 2         & 4             
\end{tabular}
\caption{The parameter settings of our 6 policies }
\vspace{-4mm}
\label{t2}
\end{table}

In summary, there are a total of 6 candidate grouping policies, and we label them as ``map2b2l'', ``map2b3l'', ``map2b4l'', ``swap2b2l'', ``swap2b3l'', ``swap2b4l''. And the benchmarks we use to evaluate these policies  cover  lots  of  functions  used in existing quantum algorithms such as QFT and algorithmic functions in classical computing. 


\subsection{Generating Group List}

The grouping process is divided into two steps. First, we partition gates into subgroups based on 2-bit constraint. Second, we partition each subgroup into smaller groups based on n-layer restriction. The two steps are illustrated in Algorithm  \ref{bit} and Algorithm  \ref{lay}, respectively.

In the bit partition step, we first transform a quantum program into a Directed Acyclic Graph (DAG). We iterate through the DAG following its topological order to ensure that a node always finds its group after its predecessor does. 
This way, we are able to greedily group a node with its parent nodes whenever possible, thereby dividing DAG into subgroups of largest possible size. 
In the layer partition step, we first label each node with its global depth, then divide nodes within each subgroup into smaller groups based on this labeled depths. 

\blue{After dividing a quantum program into groups, we ``de-duplicate'' these groups by calculating their corresponding matrices and eliminating duplicated ones. Two groups with permutated Qubits but same operations are also treated as duplicate.} 

\blue{In the static pre-compilation stage, we randomly select one-third of quantum programs from our set of benchmarks. We use these programs to generate a category of groups for the profiling purpose. The corresponding pulses of these groups will be stored and reused in future quantum programs.} 

\blue{}


\begin{algorithm}[b]
\caption{Bit Dividing}  
\label{bit}  
\begin{algorithmic} [1]
\REQUIRE qasm files, bit constraint(bc)
\ENSURE large-groups
\STATE Initialize large-groups
\FOR{qasm in all qasm files}
\STATE DAG = ToDAG(qasm)
\FOR{node in DAG.topological-order:}
\IF {node can be grouped with both predecessor}  
\STATE Merge the groups the two predecessors are in
\ELSIF{node can be grouped with one predecessor}
\STATE Group the node with the predecessor
\STATE Update large-groups
\ELSIF{node can be group with no predecessor}
\STATE Put the node in a new group
\STATE Update large-groups
\ENDIF
\ENDFOR
\ENDFOR
\RETURN large-groups
\end{algorithmic}  
\end{algorithm}

\begin{algorithm}[h!]
\caption{Layer Dividing}  
\label{lay}  
\begin{algorithmic} [1]
\REQUIRE large-groups, layer constraint(lc)
\ENSURE group list
\STATE Initialize group-list
\FOR{node in DAG.topological-order:}
\STATE Depth[node] = max(Depth[node's predecessor(s)]) + 1 
\ENDFOR
\FOR{subgroup in large-groups:}
\STATE startDepth = depth of shallowest node
\STATE layer = 0
\STATE Initialize temp-group
\FOR{node in subgroup:}
\STATE diff = depth[node] - start
\IF{diff mod lc $\leq$ layer}
\STATE Append node to temp-group
\ELSE
\STATE Append temp-group to group-list
\STATE Clear temp-group
\STATE Append node to temp-group
\STATE layer += 1
\ENDIF
\ENDFOR
\ENDFOR
\RETURN group-list
\end{algorithmic}  
\end{algorithm}


\subsection{Generating Latency List and Pulse List}
We use QOC to generate the pulse list and latency list from the category of groups. The GRAPE tool we use requires target unitary matrix, target fidelity, and target latency as inputs. The latency of a certain group is determined by a binary search. Short latency leads to more iterations with long training time and does not guarantee the convergence, while long latency loses the advantages of quantum optimal control. Therefore, binary search is necessary to ensure optimal latency within the target fidelity convergence requirement. We set the target fidelity cost function to be a typical value $1\times10^{-4}$ and maximum run time budget to be 600s for each iteration of binary search. The available methods include ADAM, BFGS, L-BFGS-B, and SLSQP. 
\blue{We choose BFGS as our optimization method for training the pulses.} To verify our idea, we use a model of a two-level spin Qubit($\omega/2\pi$: 3.9 GHz).

\subsection{Generating Overall Latency}
After generating a group list, we restructure the original DAG into a new DAG by turning each group into a node. We obtain the latency of each node by iterating through the profiling table to find its match. 

Following the topological order of the new DAG, we use dynamic programming to compute and store the until-this-step latency at each node by adding the largest latency of its predecessors to the latency of itself. The overall latency computed at the last node is the overall latency of the whole group. \blue{The detailed algorithm could be found at Algorithm  \ref{ove}}.

\begin{algorithm}[h!]
\caption{Overall Latency}  
\label{ove}  
\begin{algorithmic} [1]
\REQUIRE qasm, profile-table, latency-table
\ENSURE overall latency
\STATE Initialize latency (a list)
\FOR{node in DAG.topological-order:}
\STATE latency[node] = max(latency[node's predecessor(s)]) + latency-table[node]
\ENDFOR
\STATE overall-latency = last index of latency
\RETURN overall-latency
\end{algorithmic}  
\end{algorithm}

\subsection{The Effect of Sequence of Mapping and Grouping}

\blue{We use two methods to solve quantum hardware constraints. The first method is to first resolve the conflicts then perform the grouping operation. Such method is called as ``swap then group''. The second option is to first group these gates then concatenate the pulses of swap gates when necessary. Such method is called as ``group then swap''. We explore both options because on some quantum computer the swap operation is directly supported.}
We find that under certain circumstances, mapping then grouping has lower overall latency, in which case a swap gate is decomposed into three CNOT gates. Such method appears to have advantage most likely because those CNOT gates are more flexible in joining other gates. Furthermore, these CNOT gates are more likely to be cancelled with other CNOT gates such that the overall latency is reduced.

\subsection{Optimizing the most frequent group}


We select the group of highest frequency and spend more time training it with different methods so that the latency of this particular group could be further reduced.
The goal is to further reduce overall latency without the high overhead
when the pre-compiled 
pulses are used for a new program. 

\section{Accelerated Dynamic Compilation}
\label{at}
\subsection{The Notion of Coverage}

After the static pre-compilation, we have the pulses for 133 profiled groups.
Given a new program, it will be 
first decomposed into groups with 
\blue{``map2b4l"}
For the groups that fall into the 
pre-compiled set, we can directly use the pulses. 
We call them as ``covered'' groups and \projectname
does not incur any training overhead to generate pulses for them.
We define the coverage as follows:
$$Coverage\ Rate = \frac{\#\ Groups\ covered\ by\ our\ category}{\#\ Groups\ of\ the\ program}$$
Clearly, the coverage is an important factor determining the 
benefit of our approach. 
For the ``un-covered'' groups, we will use dynamic 
QOC based compilation to generate pulses. 
\blue{We also
call this step as training since it resembles the training process in machine learning.
The following sections will describe
our ideas to accelerate the training process.}

\begin{figure}[h!]
\centering
\includegraphics[width=0.40\textwidth]{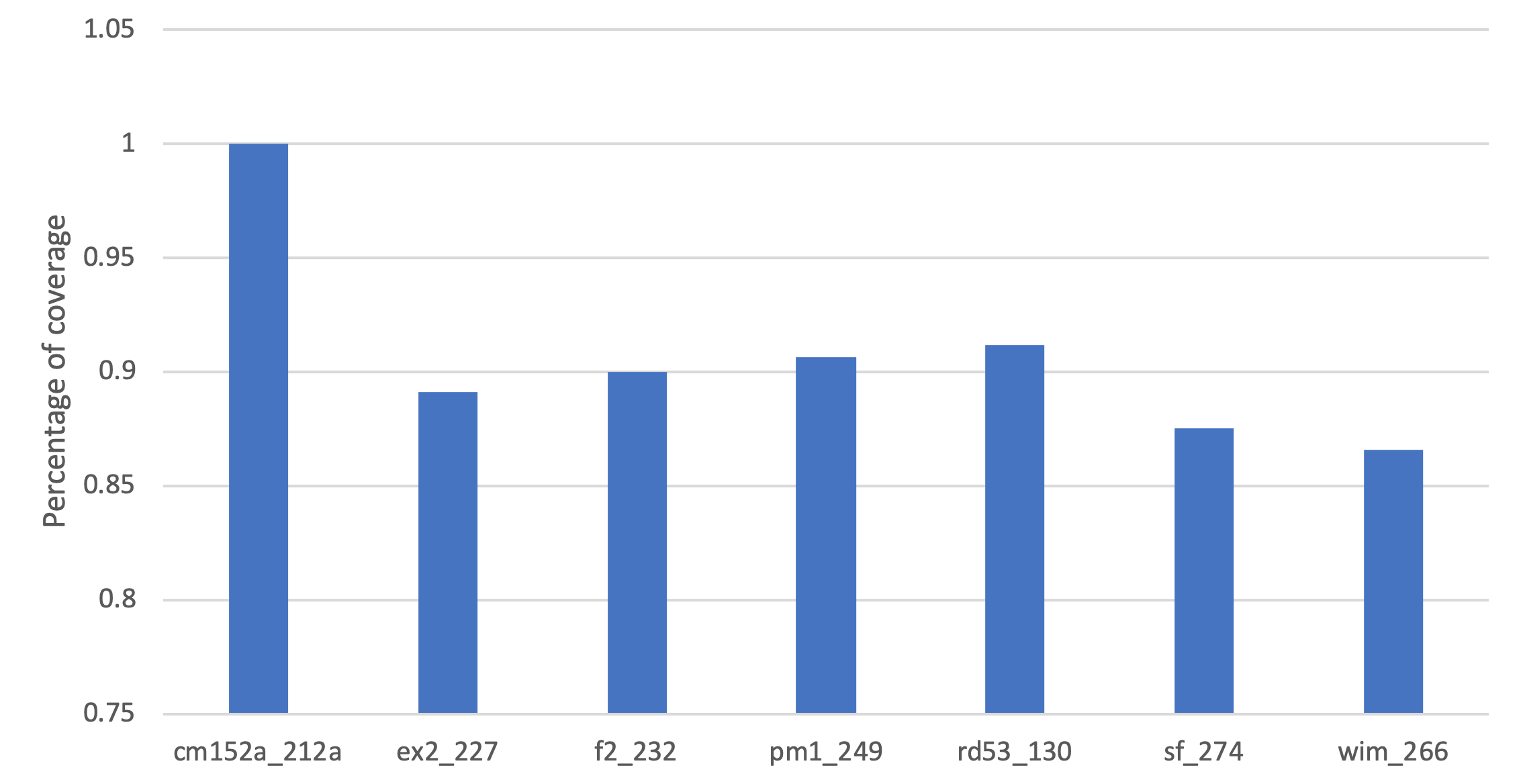}
\caption{\green{The coverage of these programs are measured under the policy of ``map2b4l''. These programs are first decomposed into groups and compared with pre-compiled groups to determine the coverage.}}
\vspace{-8mm}
\label{coverage}
\end{figure}

\subsection{Identify Similar Groups}

Since the quantum control evolves from the initial matrix to the target matrix by multiplying time-dependent, hardware-specific control Hamiltonian matrices, similar matrices could share similar pulses. \blue{Therefore, pre-computed pulses could be used}
as inputs to GRAPE to reduce the number of iterations of gradient descent when we train a new group. 


Currently we use 4 similarity functions to decide whether groups are similar. The first two are simply the difference between the matrices:
$d_1(\mathbf{A}, \mathbf{B}) = \sum_{i=1}^n \sum_{j=1}^n |a_{ij} - b_{ij}|$;
$d_2(\mathbf{A}, \mathbf{B}) = \sqrt{\sum_{i=1}^n \sum_{j=1}^n (a_{ij} - b_{ij})^2}$.
The last two similarity functions represent the fidelity of quantum unitary operations: 
$d_3(\mathbf{A}, \mathbf{B}) = Tr(A^{*}B)$;
$d_4(\mathbf{A}, \mathbf{B}) =        F(A,B)=\left(\textrm{tr}\sqrt{\sqrt{A}B\sqrt{A}}\right)^2$.


\begin{figure}[h!]
\centering
\vspace{-4mm}
\includegraphics[width=0.40\textwidth]{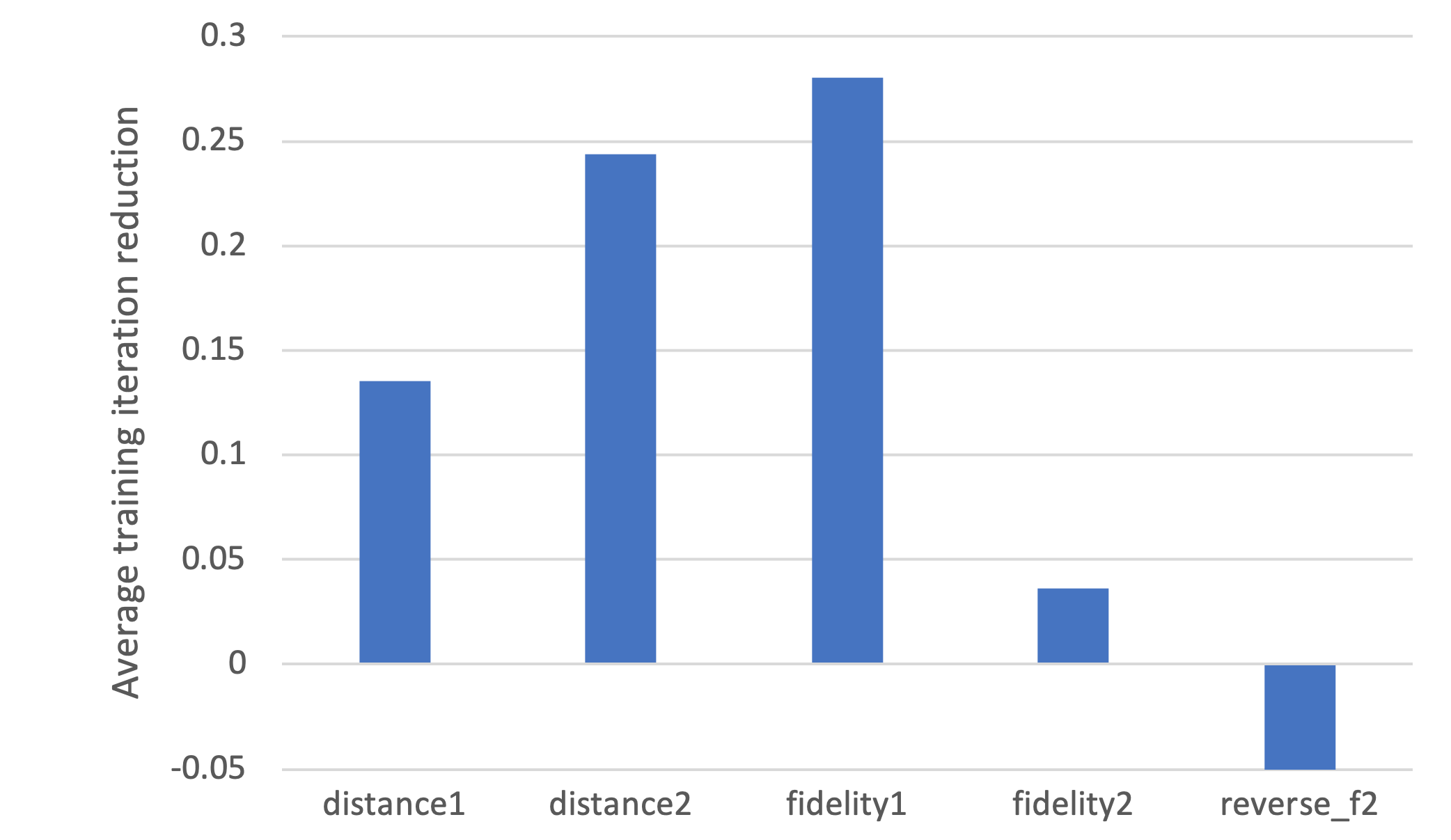}
\caption{It shows the average reduction of iterations required for computing pulses using each of the five similarity functions.}
\vspace{-2mm}
\label{8}
\end{figure}

\blue{Figure \ref{8} shows the partial results of accelerated training. We apply different similarity functions together with the SG algorithm described below. The first four similarity function attempts to measure the similarity between two groups, and the fifth one is the inverse of the fourth function and aims to measure the opposite of similarity. The result shows that ``fidelity1" function achieves most reduction of iterations, and the fifth function causes the the number of iterations to increase.}

\subsection{Determining Compiling Sequence with Similarity}

Based on the insights that the pulse of a new group can be generated faster with a similar group subject to a similarity function, 
for all un-covered groups, our goal is to generate an
ordered {\em Compilation Sequence} $CS=[g_1,g_2,...,g_n]$
(with $n$ uncovered groups)
that minimizes the estimated compile time.
According to $CS$, the pulse for $g_1$ will be generated 
first, and it will be used as the initial matrix to train
the pulse of $g_2$. 
To minimize the total compile (training) time, 
we ensure that the sum of similarity
between consecutive groups in the sequence is minimized.
With $n$ groups, we have $n!$ possible sequences (permutations),
following steps are used to obtain $CS$.

First, we can compute the similarity of {\em any} two pairs
of groups and construct a complete graph, denoted 
as {\em similarity graph (SG)}. 
In SG, each vertex is a group and the weight of edge 
is the similarity between two groups.
We also include identity matrix as a special group in SG
--- when a new group is not close enough to any groups
with pulse generated, the training of the new group will start the with 
identity matrix. 
With SG constructed,
we can compute the Minimum Spanning Tree (MST) of SG, 
which minimizes the sum of edge weights. 
In the process of generating MST using the greedy 
algorithm, i.e., Prim algorithm, we can remember the 
sequence that all vertices are selected, this sequence is 
exactly what we need for $CS$.
We choose the identity matrix as the starting point to generate
MST. 
\blue{Figure \ref{MST} shows an example of this procedure. }



\subsection{Parallelizing Compilation with Balanced MST Partition}

For a large program, the number of group can be large,
the reduction of compilation time based on group similarity 
may not be sufficient. 
The good news is that the dependency of groups in 
$CS$ is ``soft'' --- even the pulse of predecessor groups
are not available, we can always train a group starting from 
identity matrix. In fact, we can choose any order to train
the un-covered groups, the consequence is just the longer
training time. 
In this sense, we can consider that the training of each group
is independent.
Thus, the job of training all groups can be perfectly 
parallelized with multiple ``workers''~\footnote{In this paper,
we consider ``workers'' in abstract sense, and just
focus on how to partition the workload. In reality, it can be a thread, process, or a CPU/GPU, or even a node in distributed computing platform. }.

Since there is no dependency between the groups, 
the major factor determining the overall performance of 
parallel training is the workload assigned to each worker. 
In our problem formulation with SG and MST, 
the problem is how to partition the MST into multiple 
sub-graphs such that the difference of critical paths of
them is minimized.


To achieve such division, we utilize METIS~ \cite{metis},
a well-known graph partitioner to \blue{divide the MST as balanced as possible.} 
However, this balanced partitioning method operates on a graph with node weights and divides it into connected sub-parts each having a similar sum of weights. Our MST generated from SG has only weights on its edges.

\begin{figure}[t]
\begin{subfigure}{0.24\textwidth}
\vspace{-2mm}
\includegraphics[width=0.9\linewidth]{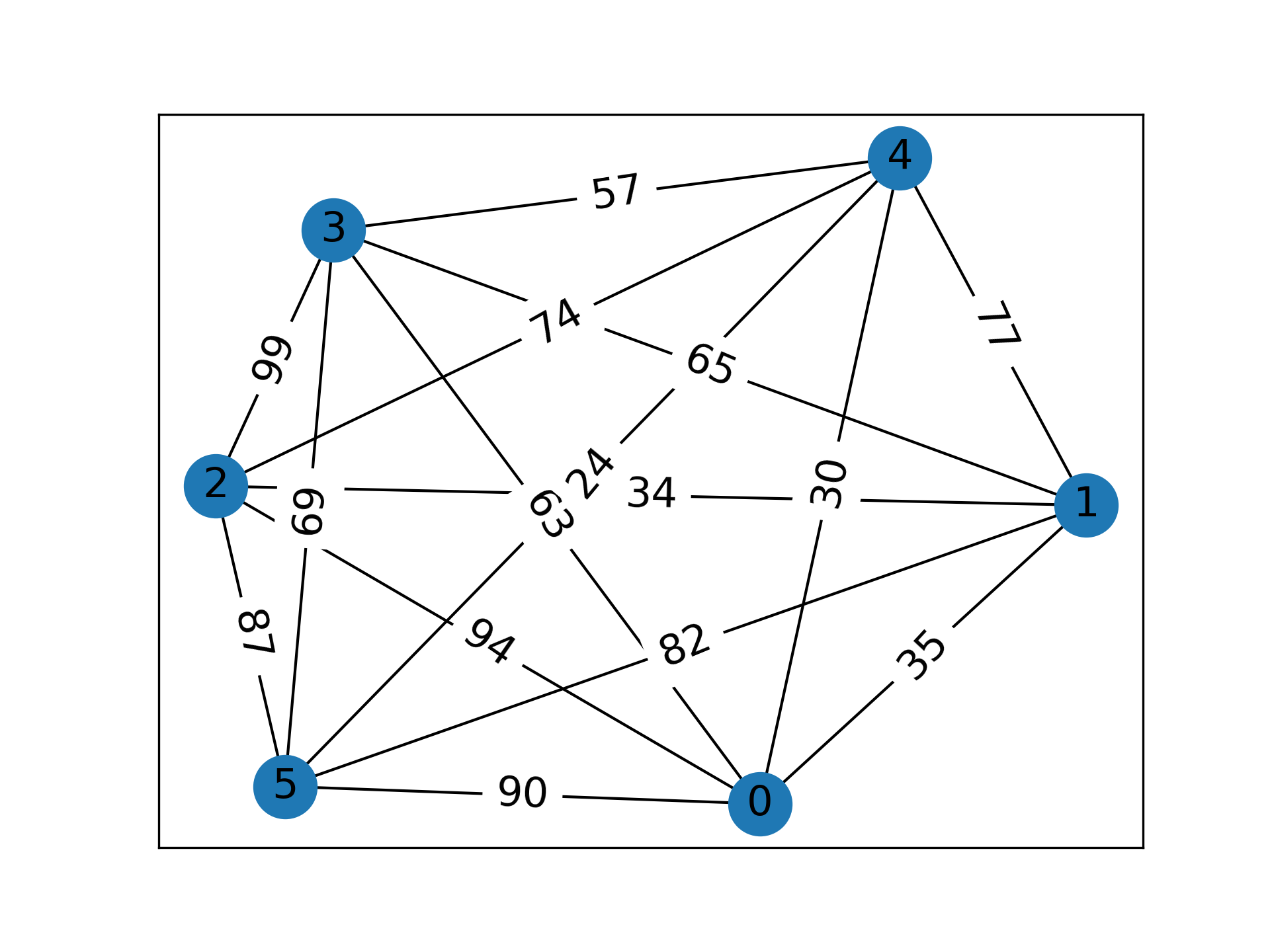} 
\vspace{-4mm}
\caption{A 6-node SG}
\label{fig:subim1}
\end{subfigure}
\begin{subfigure}{0.24\textwidth}
\includegraphics[width=0.9\linewidth]{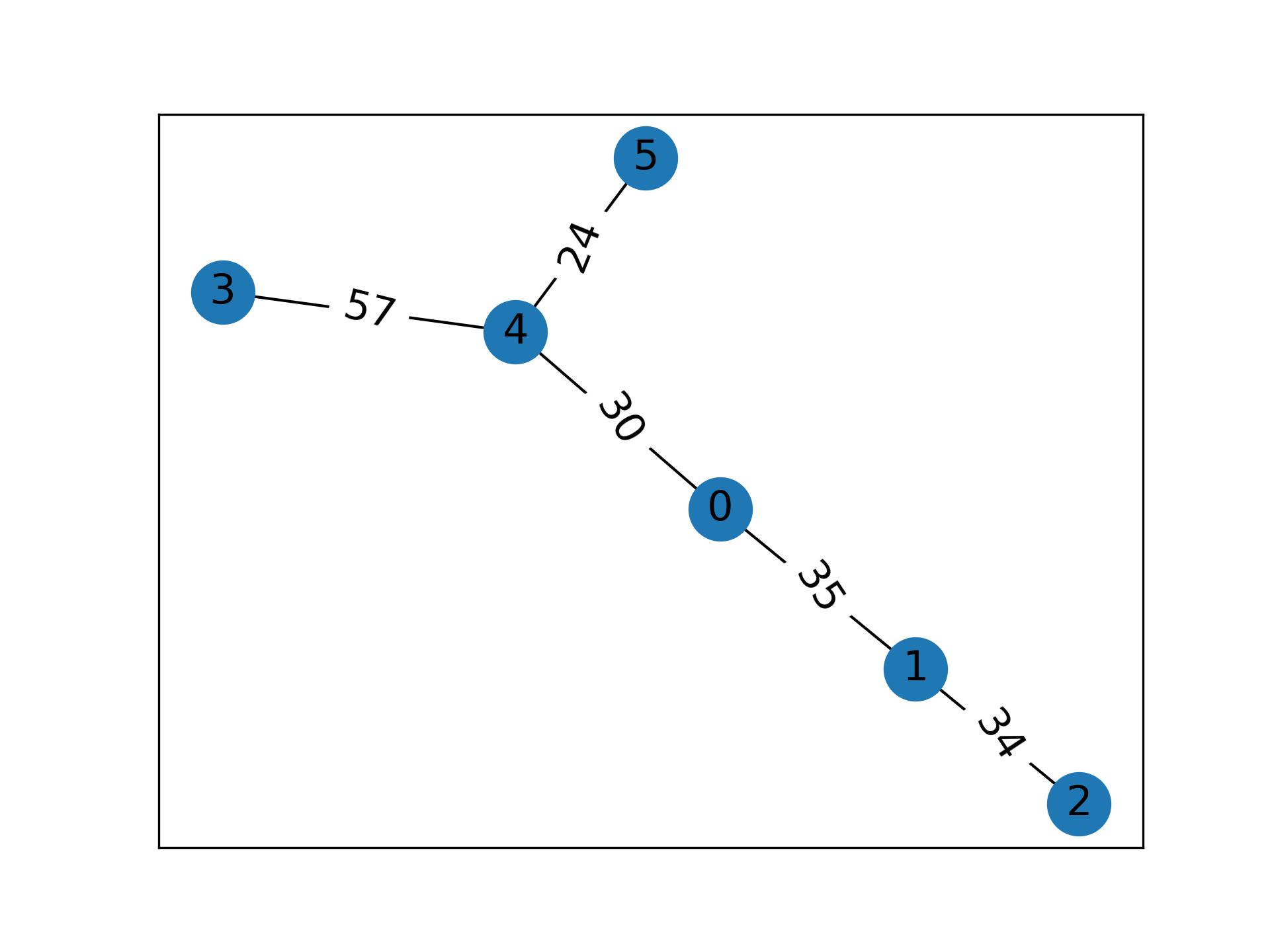}
\vspace{-4mm}
\caption{Minimum spanning tree}
\label{fig:subim2}
\end{subfigure}


\begin{subfigure}{0.24\textwidth}
\vspace{-2mm}
\includegraphics[width=0.9\linewidth]{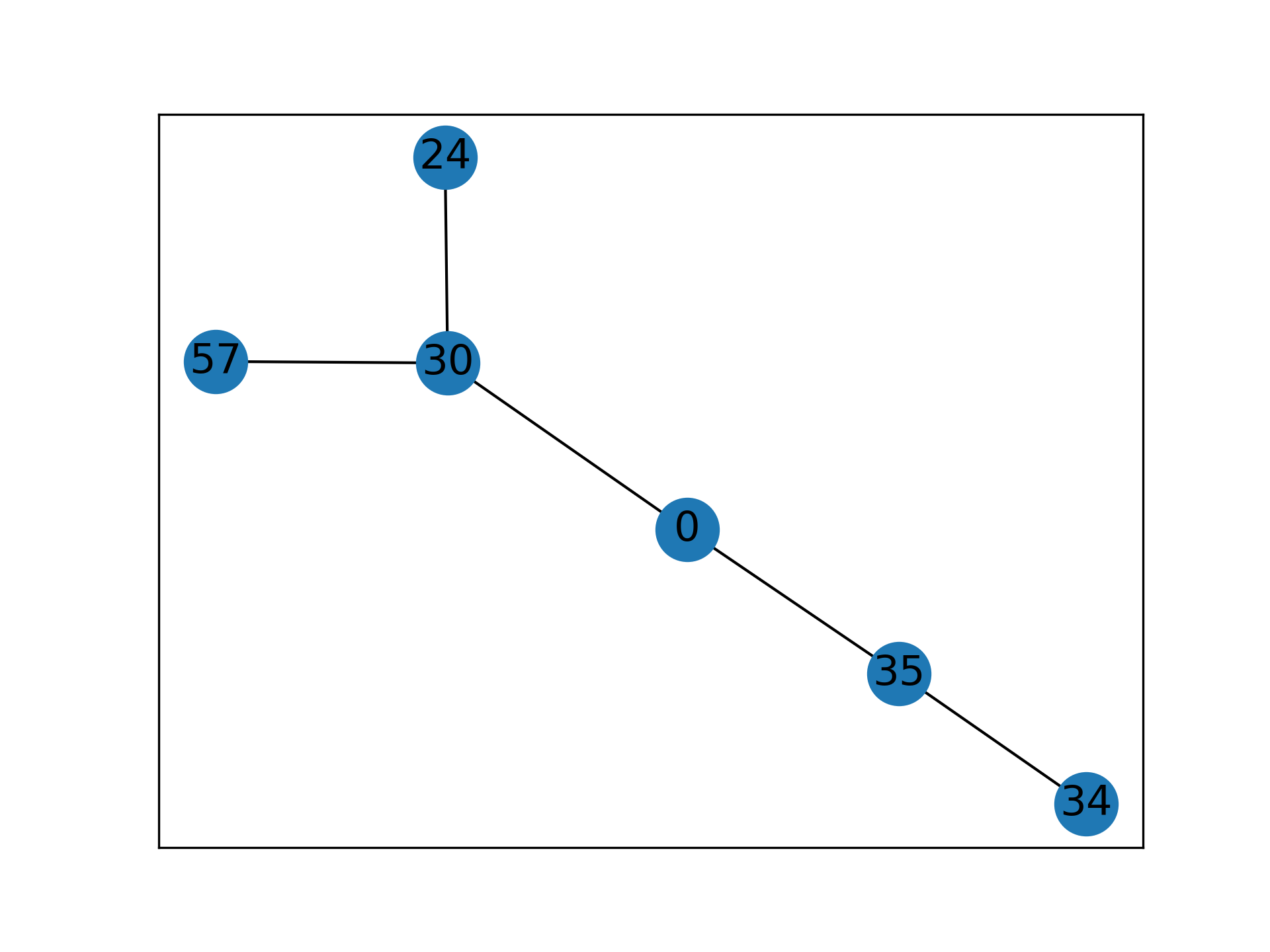}
\vspace{-4mm}
\caption{MST with shifted weight}
\label{fig:subim3}
\end{subfigure}
\begin{subfigure}{0.24\textwidth}
\includegraphics[width=0.9\linewidth]{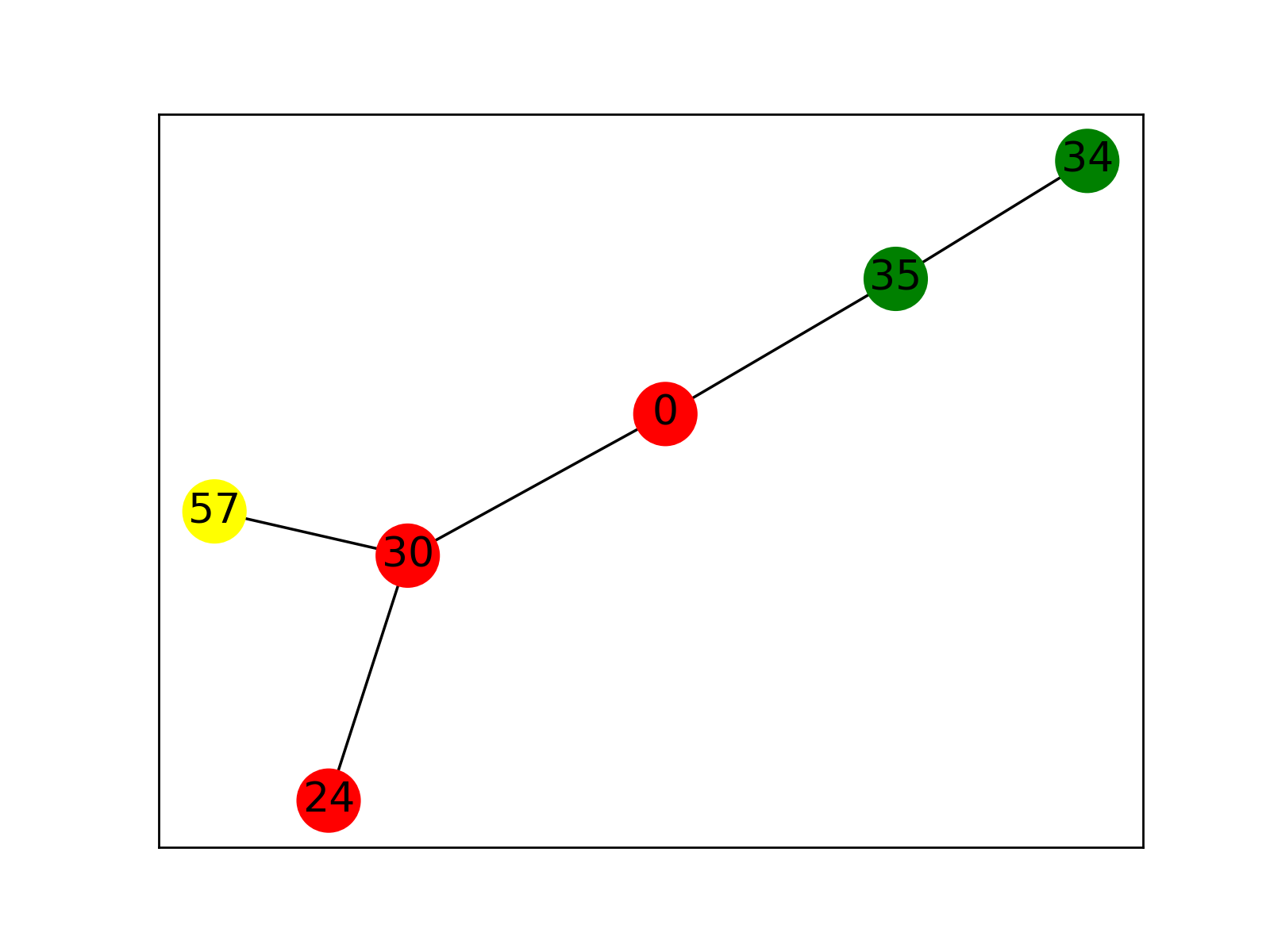}
\vspace{-4mm}
\caption{Balance partition of MST}
\label{fig:subim4}
\end{subfigure}

\caption{The procedure from SG to partitioned MST:  extract MST (b) from SG (a); shift the weight on edges to nodes (c); partition the graph into several balanced groups of similar sum of node weight, denoted by a distinct color (d).}
\vspace{-5mm}
\label{MST}
\end{figure}

To fit METIS to our problem, we transformed the MST graph with cost on edges into a new graph with weight on nodes. 
Following the optimal sequence, we shift the cost of each edge to the weight of its newly added neighboring node. The first node in the sequence is specially assigned with a value proportional to the time it takes to train the first node from 
identity matrix. \blue{This step is shown in the Figure  \ref{MST} from b to c.}

After the transformation, we use METIS \cite{metis} to divide the new graph into sub-parts, each of which is computed on a separated computer, based on its local sequence. We merge pulses of all matrices after computation is done.
\blue{Figure \ref{fig:subim4} shows an example of the partitioned graph.} 
With the generated sub-graphs, we can assign workload
to each worker to obtain the balanced execution 
during parallel processing. 


\section{Evaluation}
\label{eva}
In this section, we introduce our benchmark methodology based on the topology of IBM's 14-Qubit Quantum Computer available. Then we present our results of latency reduction and compilation time reduction.

\subsection{Benchmarks} The benchmarks are selected from previous work\cite{mapping}. 
\green{Their test circuit data, which
originated from the RevLib\cite{wille2008revlib} is available
to the public. The set of benchmarks from RevLib contains various reversible functions implemented using quantum gates. The functions include encoding functions, arithmetic functions, miscellaneous functions and symmetric functions etc. Several other functions like QFT(Quantum Fourier Transform) and GSE(Ground State Estimation)from ScaffCC\cite{javadiabhari2015scaffcc} are also in the benchmark suite.
All programs are mapped to the 14-bit IBM Q Melbourne chip~\cite{ibm_device} based on superconducting technology. } The comprehensive benchmarks cover a variety of functions used in existing quantum algorithms, \green{such as QFT in Shor algorithm}. All evaluations are conducted on a 3.4 GHz machine with 4 cores and 40 GB RAM.
\green{The whole benchmark suite includes 159 programs. We randomly sampled some quantum programs with between 200 and 2000 gates, and two QFT programs to verify our idea.
Table \ref{mix_inst} shows the instruction mixes of 6 programs and 
the average of all programs. }

\begin{table}[]
\centering
\begin{tabular}{c|c|c|c|c|c|c}
            & x      & t    & h    & cx   & rz    & tdg  \\ \hline
4gt4-v0\_79 & 0      & 56   & 28   & 105  & 0     & 42   \\ \hline
cm152a\_212 & 5      & 304  & 152  & 532  & 0     & 228  \\ \hline
qft\_10     & 0      & 0    & 20   & 90   & 90    & 0    \\ \hline
qft\_16     & 0      & 0    & 32   & 240  & 240   & 0    \\ \hline
ex2\_227    & 5      & 156  & 78   & 275  & 0     & 117  \\ \hline
f2\_232     & 6      & 300  & 150  & 525  & 0     & 225  \\ \hline
all         & 0.10\% & 22\% & 15\% & 45\% & 1.1\% & 17\%
\end{tabular}
\caption{Instruction Mixes of Benchmark Programs}
\vspace{-4mm}
\label{mix_inst}
\end{table}

\subsection{Hardware Model} 
\begin{figure}[h!]
\centering
\vspace{-4mm}
\includegraphics[width=0.40\textwidth]{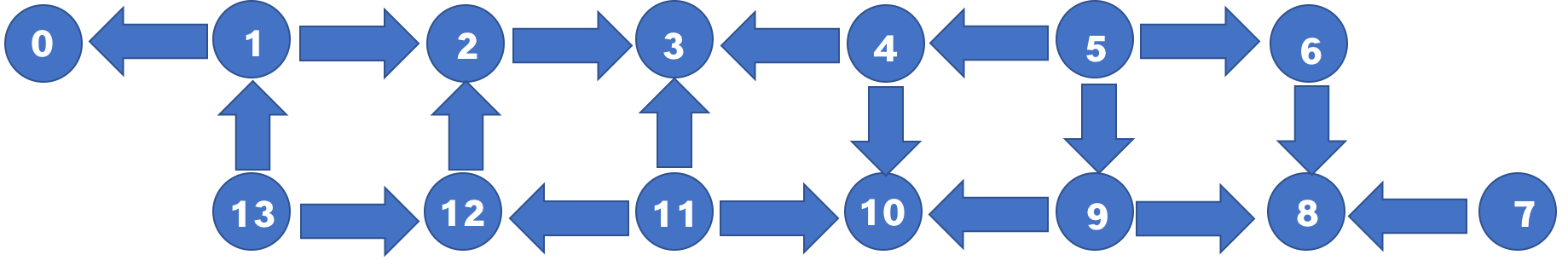}
\caption{Qubits connection of IBM's melbourne chip. \cite{qiskit}}
\vspace{-4mm}
\label{tpo}
\end{figure}

We use the topology from IBM Q Melbourne chip where the two-Qubit gates are not symmetric and therefore have certain directions. CNOT gate is only allowed in one direction. We extend the mapping algorithm of  \cite{mapping} with consideration of cross-talk to insert swap gates into the program. The direction of the 2-bit gate and topology of Melbourne is showed in Figure  \ref{tpo}.

\subsection{Mitigation of Cross-talk in Mapping}
\label{cross}
\blue{We quantify the overall cross-talk effect as the total occurrences of close CNOT operations in each layer. \green{This metric is adopted from the paper\cite{murali2020software}. 
The rational to use this metric is discussed in 
Section~\ref{crosstalk}}. Before and after our mapping algorithm, the cross-talk effect deceases for most of the tested quantum programs and we observe an average of 17.6\% reduction of cross-talk effect.} 
The systematic method to mitigate cross-talk 
effect is still an open question. 
Nevertheless, our workflow is the among the first to tackle the problem of cross-talk effect by Qubit mapping.
We leave the systematic study of the problem for 
future work. 

\subsection{Reduction of Overall Latency}

\begin{figure}[h!]
\centering
\includegraphics[width=0.50\textwidth]{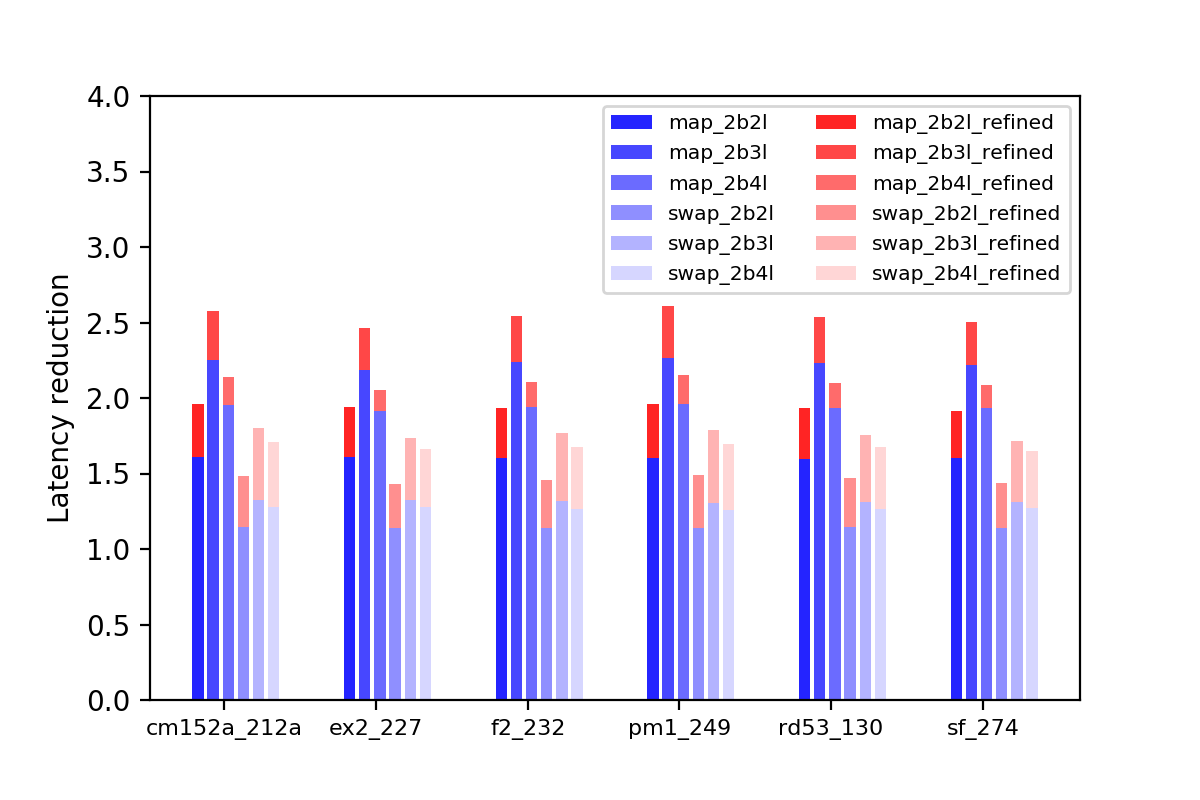}
\vspace{-6mm}
\caption {\green{The latency reduction for 6 quantum programs is shown, each with 6 policies differentiated by the gradually-fading color scheme. The red ones represent the case in which the most frequent group is targeted for optimization, compared against blue ones that omit this step.}}
\vspace{-4mm}
\label{latency}
\end{figure}

\begin{figure}[h!]
 \centering
\includegraphics[width=0.50\textwidth]{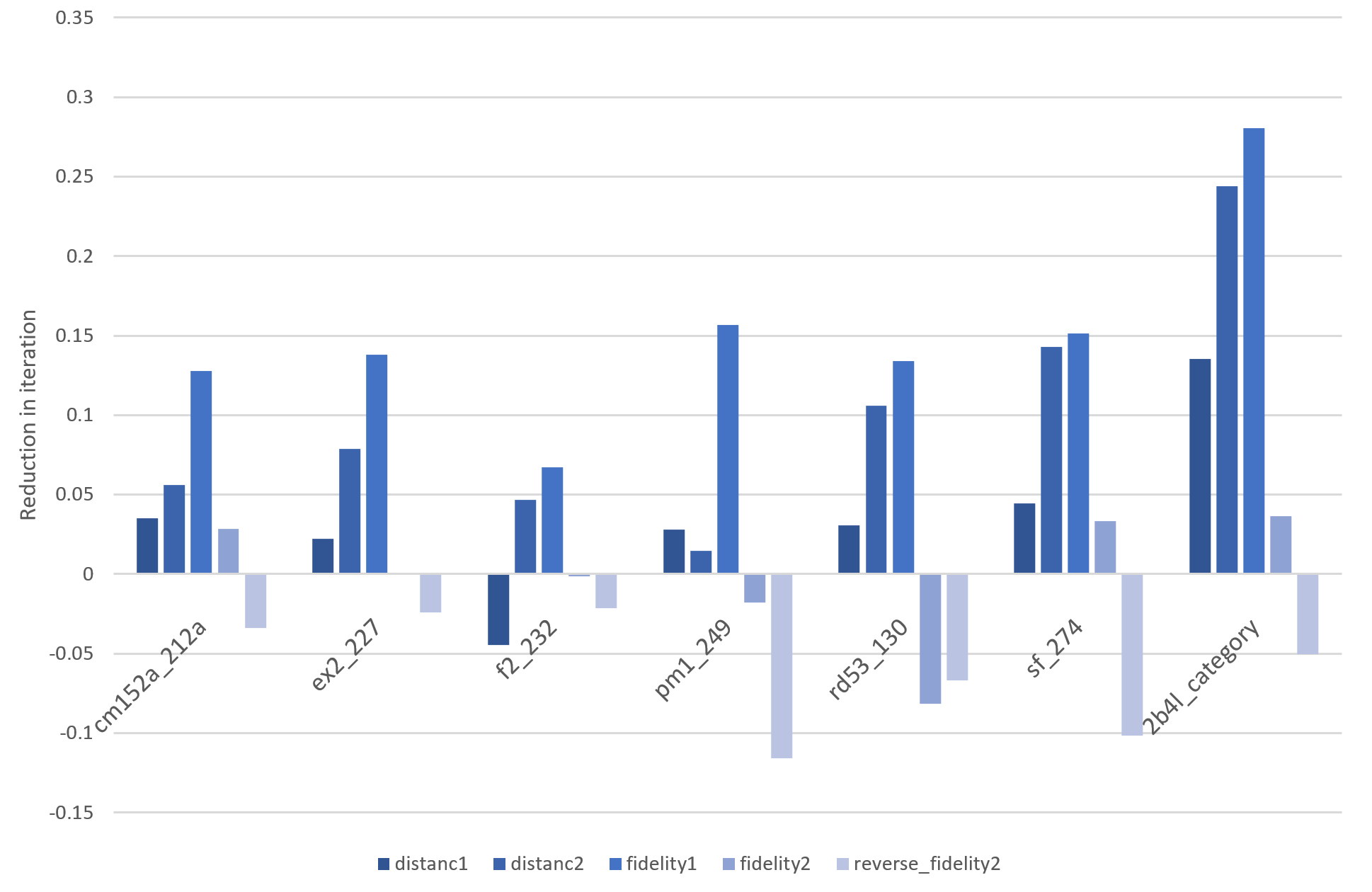}
\vspace{-2mm}
\caption{Reduction of training iteration: for each of the 7 quantum programs (including a set of profiled groups by map2b4l), 5 similarity functions are applied. }
\vspace{-2mm}
\label{iterations}
\end{figure}
We have implemented 6 different grouping policies. These policies feature different sizes of groups. The latency reduction of 6 grouping policies mostly lies between 1.2$\times$ and 2.6$\times$ compared with gate-based compilation. 

In our static pre-compilation process, a target program will have groups of gates falling into the profiled category. Among these groups, we pick the one with highest frequency and re-generate the pulse for this group. We will spend more computational resource training this group such that its latency could be reduced. This brings further latency reduction as plotted in the Figure \ref{latency}.

\subsection{The effect of the sequence of mapping and grouping}
The sequence of mapping and pulse generation in the control flow is also considered. The mapping algorithm of \cite{mapping} is adopted. We find that under certain circumstances, ``mapping then grouping'' has lower overall latency, in which case the swap gate is decomposed into three CNOT gates. These CNOT gates could form a more flexible group and are more likely to get diminished such that the overall latency could be reduced. However, when swap operations are directly supported 
in the target machine, it is better to choose the policy ``grouping then mapping'', since hardware-supported operations are often optimized to reach high fidelity with low latency.

\subsection{Coverage}

Here we show the coverage of 7 quantum programs under the ``map2b4l" policy. Since the category of groups is profiled by one third of our benchmarks. The profiled category is likely to contain most of groups that will appear in other programs. We achieve an average coverage of \blue{89.7\%}. 

Moreover, Figure \ref{groups} shows that the number of 2b4l groups grows much slower than linearly(though not strictly logarithmic) with the number of gates, meaning the probability of encountering uncovered groups does not scale with the size of a quantum program. It demonstrates the high reusability of pre-compiled groups.

Higher coverage corresponds to lower compilation time, since pulses of covered groups are pre-computed.
However, with larger quantum programs in the future, we expect to see lower coverage but more latency reduction for each group. When that happens, MST-accelerated QOC will come to use and demonstrate its powerfulness.
Hence, there exists a clear trade-off to be considered between the pre-compilation overhead and coverage, especially when dealing with large quantum programs.

\begin{figure}[h!]
\begin{subfigure}{0.24\textwidth}
\centering
\includegraphics[width=0.9\linewidth]{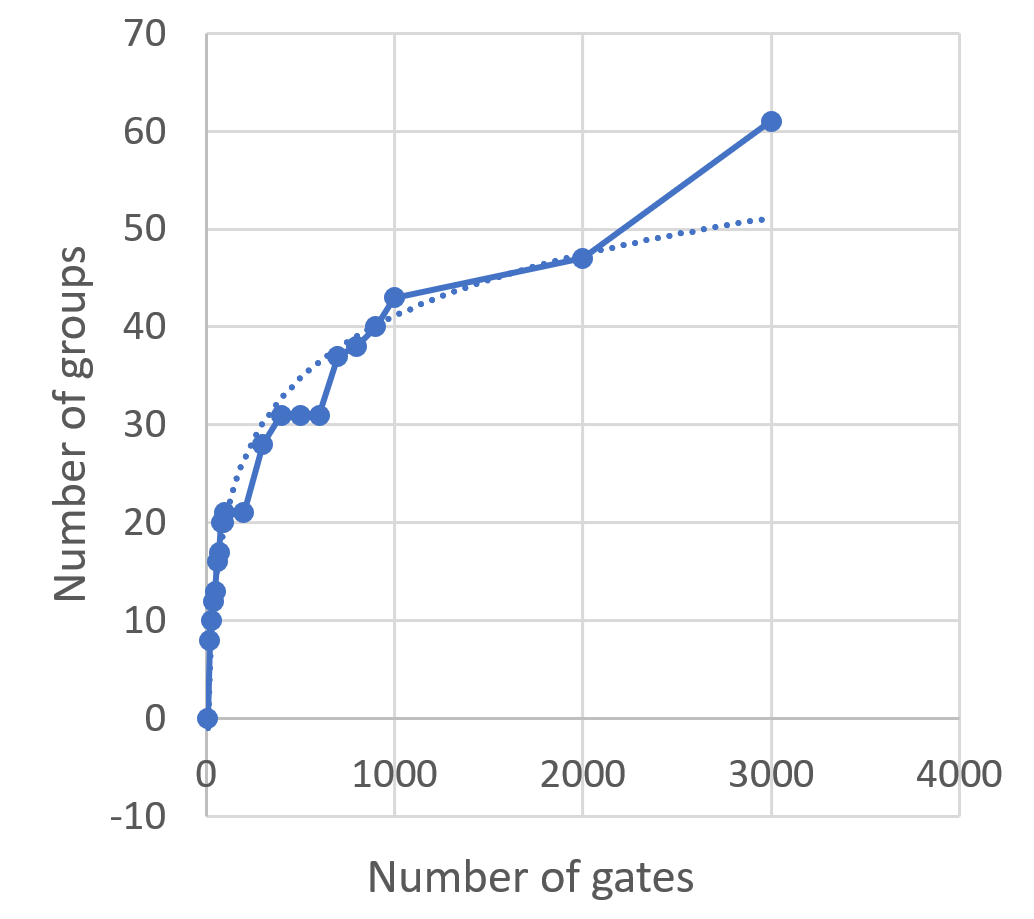} 
\caption{\# groups vs \# gates}
\label{fig:subim1}
\end{subfigure}
\begin{subfigure}{0.24\textwidth}
\includegraphics[width=0.9\linewidth]{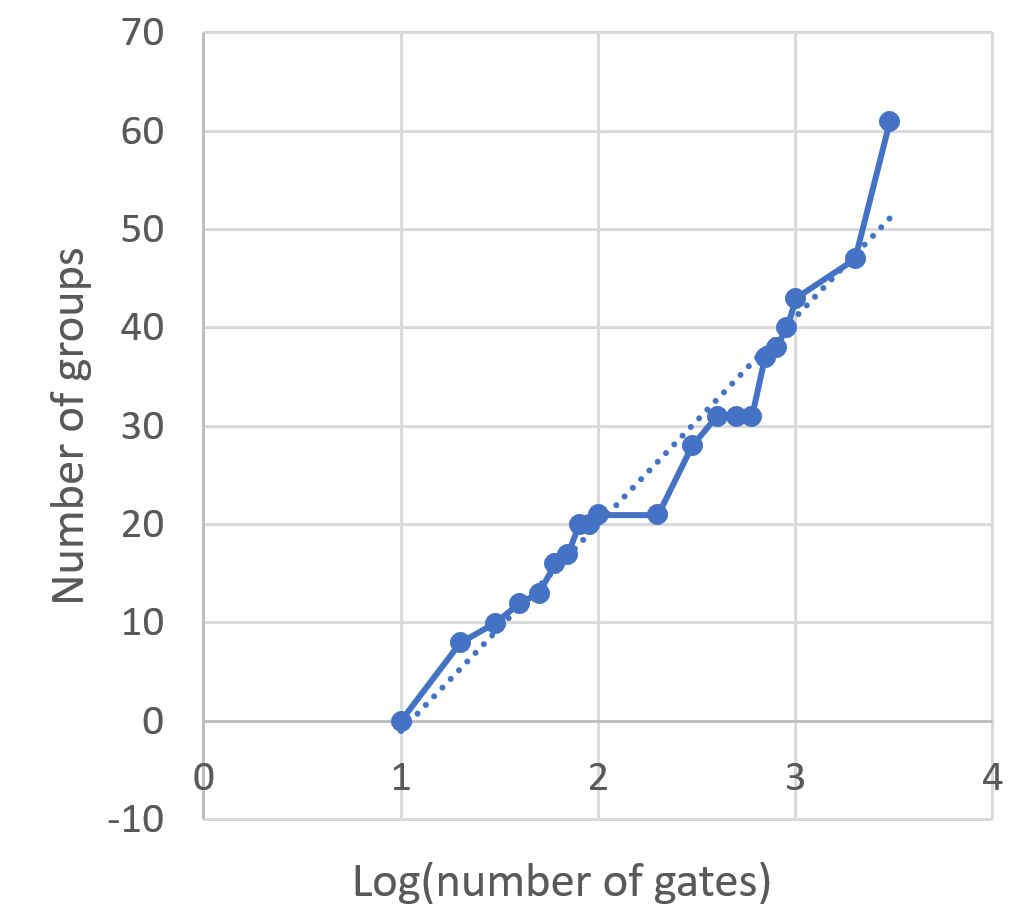}
\caption{\# groups vs log(\# gates)}
\label{fig:subim2}
\end{subfigure}
 
\caption{The figure demonstrates the relationship between the growth of group number and that of gate number.}
\vspace{-4mm}
\label{groups}
\end{figure}

\subsection{Accelerated Training}

In this part, we compare the training iterations of groups with and without accelerated training. To directly show the better performance of accelerated training, we demonstrate this methodology with a pre-compiled category under the ``map2b4l" policy, which has 133 groups. As shown in Figure \ref{iterations}, we could reach a max iteration reduction of \blue{28\%}. The acceleration highly relies on the size of MST we have. For a larger MST, the two group connected are more likely to be very close to each other. We expect to see more iteration reduction with larger quantum programs. \green{We use iteration reduction as our metric for computing source because the running time of optimal control grows linearly with the number of iterations. 
In our experiments, multiple optimal control programs are executed 
in a multi-programmed fashion. We cannot ensure each execution 
get equal system resource. Therefore, when measuring the 
improvement of \projectname, it is more reasonable 
and accurate to report the number of iterations,
instead of running time.}

\begin{figure}[t]
\centering
\includegraphics[width=0.40\textwidth]{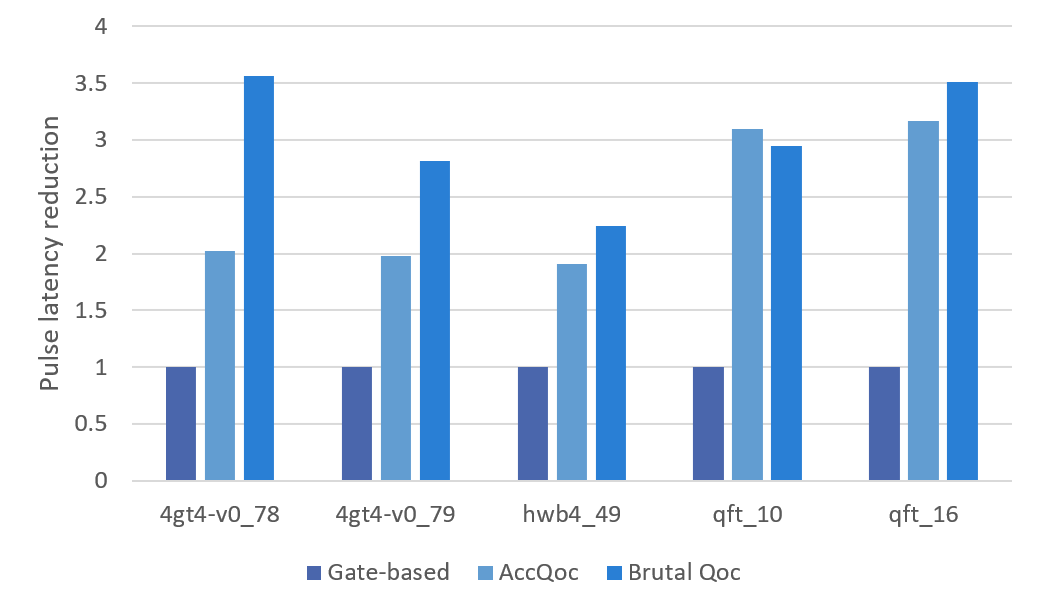}
\caption{\green{Comparing \projectname with Brute-Force QOC Training}}
\vspace{-6mm}
\label{brutal}
\end{figure}

\subsection{Compilation Time}

The most time-consuming part of our workflow is the step of generating pulse list and latency list. After the pre-compilation is finished, it takes no time to look up the pulses for covered groups. 
Quantum programs with higher coverage will gain more compilation time reduction, since less groups need to be compiled to pulses. 
Moreover, since the size of our pre-compiled category does not scale linearly with the size of quantum programs, larger quantum programs will feature more compilation reduction.

\green{To show the relation between compile time reduction and
latency reduction, we choose the ``brute force'' QOC's compilation time as our baseline. The same methodology is also used in \cite{gokhale2019partial}. We form the ``brute force QOC'' groups by including as many qubits and gates as possible. The Figure \ref{brutal} shows that the average latency reduction of {\em \projectname} is 2.43$\times$ while the ``brute force QOC'' achieves 3.01$\times$ reduction. \cite{gokhale2019partial} achieves similar 
latency reduction as \projectname. 
On the other side, the compilation time reduction 
compared to ``brute force'' QOC compilation is 9.88$\times$.
Thus, our method trades off minor latency reduction (2.43$\times$ compared with 3.01$\times$) for significant compile time
reduction (9.88$\times$).}


\section{Related Work}
\label{re}
The traditional gate-based workflow of quantum computing has been well studied \cite{larose2019overview,murali2019full,preskill2018quantum}. Techniques have been proposed to optimize the compiler frontend \cite{javadiabhari2015scaffcc} and the mapping problem \cite{li2019tackling,mapping}. After the idea of quantum optimal control is proposed, some research has focused on the concrete implementation of  algorithms \cite{de2011second,khaneja2005optimal,knight2017ibm,sklarz2002loading,zhu1998rapidly}. The optimal control tool developed by  \cite{leung2017speedup} uses GPU-based automatic differentiation to accelerate the iterations when input matrix evolving towards target matrix. Recent work \cite{shi2019optimized} noticed the problem of large overhead of quantum optimal even with GPU acceleration. \green{However, the methodology used in the paper does not address this problem and their methodology does not scale with the size of the quantum programs. The idea of utilizing commutativity could increase parallelism among groups; but those groups are relatively large
to show benefits, making it slow for quantum optimal control to compute. For example, the paper shows that the aggregated gates form  groups that contain up to 10 qubits, the QOC of such groups could take hours to generate the pulses.} 
Our solution will not require much computation resource once our category of groups is pre-compiled.
\green{The prior work \cite{kudrow2013quantum} builds a library of gates with specific rotation angles. }
A more recent work \cite{gokhale2019partial} applies idea of pre-compilation on Quantum Variational Algorithms. \green{The proposed methodology of hyperparameter tuning works well with the iterative execution of variational algorithms. However, the methodology cannot be applied to the static groups in non-variational algorithms. 
Compared to them, \projectname is more general and not limited to any special types of quantum algorithms. The hybrid characteristic of our method makes it suitable for both variational algorithms like VQE and non-variational algorithms like Shor algorithm.
It can support arbitrary rotation angles since it simply corresponds
to a different matrix.} 

Most importantly, our workflow does not scale linearly with the size of input program. Therefore, our workflow {\em \projectname} is more scalable and more suitable for future compilation of large quantum programs.

\section{Conclusion}
\label{con}

In this paper, we propose {\em \projectname}, a comprehensive 
static/dynamic hybrid workflow to transform gate groups
(equivalent to matrices) to pulses using 
QOC (Quantum Optimal Control) with a reasonable compilation time
budget. \projectname is composed of static pre-compilation
and accelerated dynamic compilation. 
We leverage static pre-compilation to generate 
pulses for the frequently used groups to eliminate the 
dynamic compilation time for them.
The pulse is generated using QOC with binary search to 
determine the latency. 
For a new program, the dynamic compilation deals with ``un-covered'' groups
with accelerated pulse generation.
The key insight is that the pulse of a group 
can be generated faster based on the generated pulse
of a {\em similar} group.
We propose to reduce the compilation time by generating 
an ordered sequence of groups in which the sum of similarity
among consecutive groups in the sequence is minimized. 
\blue{With the methodology of \projectname, we reached a balanced point of compilation time and overall latency. The results show that accelerated compilation based on MST \green{achieves ${9.88}\times$ compilation speedup compared to the standard compilation of each group} while maintaining an average $2.43\times$ latency reduction compared with gate-based compilation.}


\bibliographystyle{plain}
\bibliography{refs}

\end{document}